**Towards a comprehensive understanding of the low energy luminescence peak in 2D materials**

Keerthana S Kumar[1], Ajit Kumar Dash[1], Hasna Sabreen H[2], Manvi Verma[1], Vivek Kumar[1], Kenji Watanabe[3], Takashi Taniguchi[4], Gopalakrishnan Sai Gautam[2], Akshay Singh[1, *]

[1]Department of Physics, Indian Institute of Science, Bengaluru, Karnataka -560012, India

[2]Department of Materials Engineering, Indian Institute of Science, Bengaluru, Karnataka -560012, India

[3]Research Center for Functional Materials, National Institute for Materials Science, Ibaraki 305-0044, Japan

[4]International Center for Materials Nanoarchitectonics, National Institute for Materials Science, Ibaraki 305-0044, Japan

*Corresponding author: aksy@iisc.ac.in

**Abstract**:

An intense low-energy broad luminescence peak (L-peak) is usually observed in 2D transition metal dichalcogenides (TMDs) at low temperatures. L-peak has earlier been attributed to bound excitons, but its origins are widely debated with direct consequences on optoelectronic properties. To decouple the contributions of physisorbed and chemisorbed oxygen, organic adsorbates, and strain on L-peak, we measured a series of monolayer (ML) $MoS_2$ samples (mechanically exfoliated (ME), synthesized by oxygen-assisted chemical vapour deposition (O-CVD), hexagonal boron nitride (hBN) covered and hBN encapsulated). Emergence of L-peak below 150 K and saturation of photoluminescence (PL) intensity with laser power confirm bound nature of L-peak. Anomalously at room temperature, O-CVD samples show high A-exciton PL (c.f. ME), but reduced PL at low temperatures, which is attributed to strain-induced direct-to-indirect bandgap change in low defect O-CVD $MoS_2$. Further, L-peak redshifts dramatically ~ 130 meV for O-CVD samples (c.f. ME). These observations are fully consistent with our predictions from density functional theory (DFT) calculations, considering effects of both strain and defects, and supported by Raman spectroscopy. In ME samples, charged oxygen adatoms are identified as thermodynamically favourable defects which can create in-gap states, and contribute to the L-peak. The useful effect of hBN is found to originate from reduction of charged oxygen adatoms and hydrocarbon complexes. This combined experimental-theoretical study allows an enriched understanding of L-peak and beneficial impact of hBN, and motivates collective studies of strain and defects with direct impact on optoelectronics and quantum technologies.

**Introduction:**

Defects in two-dimensional (2D) materials are of 0D (vacancy, interstitial defect complexes) and 1D (line defects, grain boundaries) nature.[1–3] 0D defects are especially attractive for photon emitting applications, including possibilities as photon emitters, as well as understanding dipolar interactions in the case of closely spaced defects. Defects in $MoS_2$ and other 2D materials created using various methods including electron beam irradiation[4], strain (via nanopillars), or annealing in gaseous environment[5], have been well studied using optical and optoelectronic methods.[6] For example, broad low energy peaks have been found in exfoliated monolayer (ML) $MoS_2$ in low temperature (LT) photoluminescence (PL) spectroscopy, and are attributed in studies to either adsorbates or sulphur vacancies.[7] In other 2D materials like $WSe_2$, a series of low energy peaks have been observed, attributed to defects and dark excitons.[8]

The major issue of comparing between different reports is the lack of uniformity of starting material and process control. Importantly, defect nature/density and strain are incredibly hard to compare between studies, making it difficult to reconcile with calculations of thermodynamic defect formation energies and optical emission energies.[9–12] The substrate and environment also make a significant difference, especially in terms of background doping.[9,13] To fully understand defects in 2D materials, a range of defect densities need to be studied. Secondly, for understanding effect of strain, synthesis method dependence (chemical vapor deposition, CVD, and mechanical exfoliation, ME), as well as processing method dependence (encapsulation, covering) need to be understood in a comprehensive manner. Further, a recent work discusses the physical reason behind usefulness of encapsulation, in terms of oxygen passivation of chalcogen vacancies.[14] Thus, the complex dependence of oxygen in terms of physical and chemical adsorption is also important to understand.

We provide a comprehensive experimental-theoretical framework for the study of the low-energy PL peak in $MoS_2$ synthesized using two methods: ME and oxygen assisted CVD (O-CVD). In both synthesised versions, we modify the dielectric environment and defect density using hBN covering and hBN encapsulation. We are thus able to change the chemico-physical environment, create different kinds of defects, and modify strain. To understand the physical origin of peaks, we study these samples using optical methods (PL and Raman spectroscopy) at room temperature and cryogenic temperature (4 K), as well as perform power dependence of PL intensity. We find anomalously high PL of O-CVD samples at room temperature (c.f. ME $MoS_2$), but low A-exciton PL intensity at cryogenic temperature.

A rich variety of defects in $MoS_2$ are observed, as evidenced by varying luminescence peaks, and large shift of L-peak in O-CVD samples. Complementary Raman measurements are able to distinguish between strain (or defects) and doping, and indicate modified screening in hBN-modified samples, and increased strain for O-CVD samples. We also probe the surface composition of these samples using X-ray photoelectron spectroscopy (XPS), thus understanding the complex effects of oxygen. Detailed density functional theory (DFT) calculations on defect formation energies (for stability), and band-structure calculations incorporating strain and defects are performed. We explain the measurements on the basis of defects and strain-induced change of the nature of the bandgap. We find that L-peak in ME samples originates from a combination of charged O adatoms, sulphur vacancies, and hydrocarbon complexes, whereas O adatoms do not contribute to the case of O-CVD samples. We are thus able to provide a comprehensive understanding of the interplay of defects, oxygen (environmental) and dielectric environment, as well as strain. The beneficial impact of hBN encapsulation is attributed to reduction in charged O adatoms and hydrocarbon complexes. We find that effect of strain and defects need to be considered together in understanding these 2D materials.

**Results and Discussions**

To understand the rich variety of defects and effect of strain in $MoS_2$, we prepared a series of ML samples (summarized in Table 1) with varying defect densities and processing conditions. The CVD synthesis process uses a small amount of oxygen along with the carrier gas (see Methods), reasoned to reduce nucleation density on substrate, and prevent metal oxide precursor poisoning.[15,16] The CVD samples grown on $SiO_2$/Si are expected to be tensile strained due to different thermal expansion coefficients of the substrate and $MoS_2$.[9,13,17]

| Sample Label | Sample details |
| --- | --- |
| S1 | Exfoliated bare $MoS_2$ monolayer (ML) |
| S2 | Exfoliated ML $MoS_2$ with hBN covering |
| S3 | Exfoliated ML $MoS_2$ encapsulated between hBN layers |
| S4 | O-CVD grown ML $MoS_2$ on $SiO_2$/Si |
| S5 | O-CVD grown ML $MoS_2$ with hBN covering |

**Table 1**. Details of the samples considered in this work.

Figure 1a indicates the impact of environment (adsorbates, oxygen) on a typical 2D material, along with a sulphur vacancy ($V_S$, i.e., the defect with the lowest formation energy). A typical image of a ML grown through O-CVD is indicated in Figure 1b, showing uniform optical contrast. Images of other samples can be found in the Supplementary Information SI-I. Layer thickness is confirmed to be that of a ML using RAW optical contrast[18] and PL.

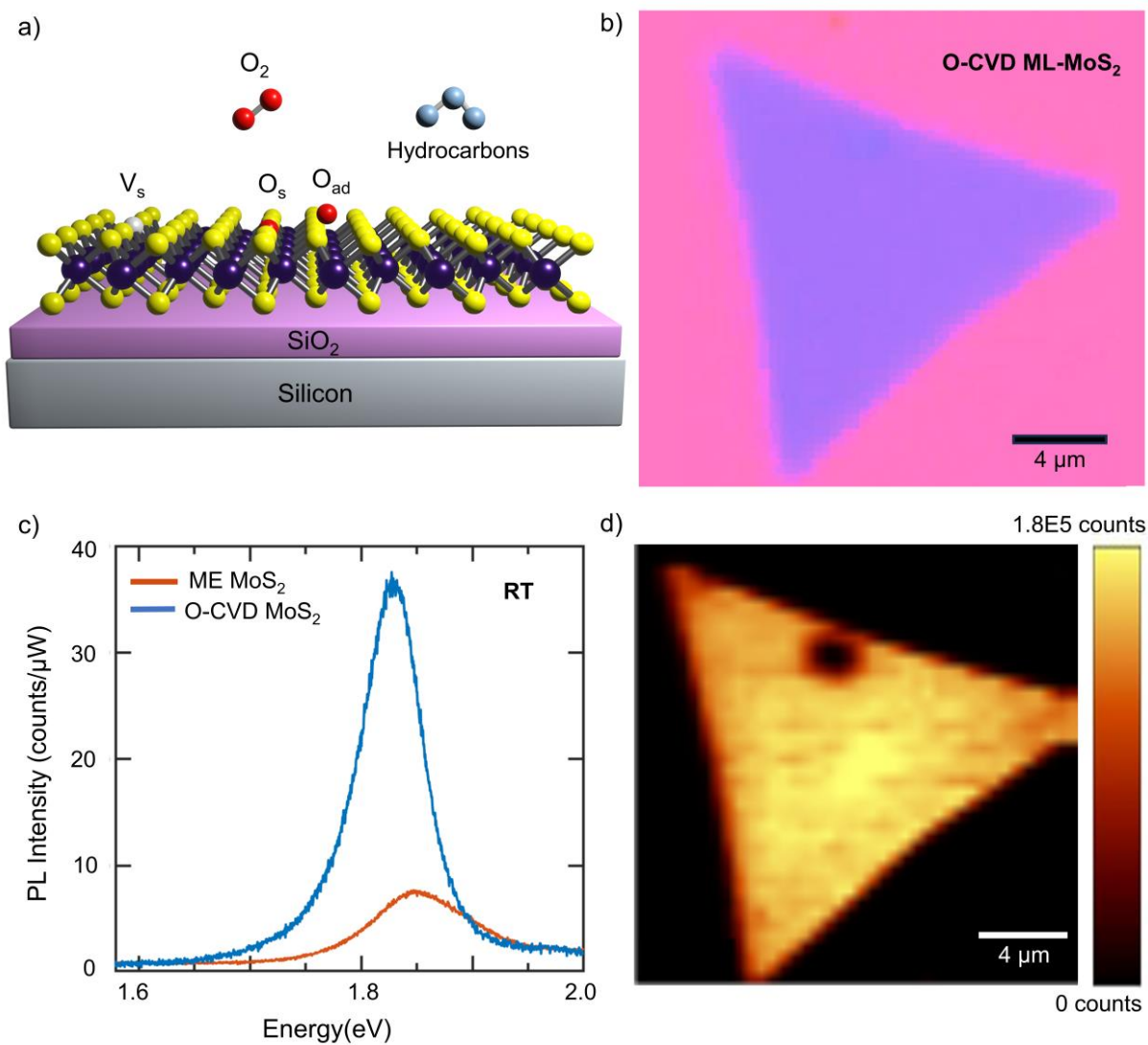

**Figure 1.** a) Schematic of monolayer (ML)-MoS$_2$ showing sulphur vacancies (V$_S$), oxygen on sulphur anti-sites (O$_S$), and adsorbates (oxygen adatom O$_{ad}$, and hydrocarbons). b) Optical microscope image of ML-MoS$_2$ synthesized using oxygen assisted chemical vapor deposition (O-CVD). c) Comparison of room temperature photoluminescence (RT-PL) in ML-MoS$_2$ obtained using O-CVD and mechanical exfoliation (ME). d) PL mapping of O-CVD ML-MoS$_2$, demonstrating nearly homogenous luminescence. The small area with low PL intensity is a bulk particle.

Firstly, we measure the PL of O-CVD (sample S4) and bare ME (S1) samples at room temperature (296 K, RT). Interestingly, we observe higher PL intensity for S4 in comparison with S1 (Figure 1c), by 50-300% (statistical data on various O-CVD samples is provided in Supplementary Figure S-XIII). We also note that there is a ~ 40 meV shift of the A-exciton peak for S4 (compared to S1), which can be attributed to both synthesis-induced biaxial tensile strain, and defect-induced doping. The observation of high PL is in contrast with usual expectations, wherein higher density of defects in CVD-grown $MoS_2$ are anticipated due to the high temperature used during synthesis. The presence of defects can induce in-gap states that can trap carriers (electrons or holes) and lead to non-radiative channels.[19,20] We also perform RT-PL mapping of S4 (Figure 1d), and observe uniform integrated PL intensity, indicating lack of inhomogeneities (non-uniform strain or defects) on the sample. Thus at RT, we can conclude that O-CVD samples will generally yield higher PL intensities than ME samples. For unoptimized growth, however, PL intensity may be reduced for O-CVD samples as well. We also note that S1 is expected to have a sizeable density of native defects and is not a pristine sample.

Higher PL intensity for O-CVD samples can be related to a combination of passivation of defects, higher quality of synthesis and/or strain. We discuss these mechanisms in sequence. Defect sites are active sites for physisorption of adsorbates (e.g., organic molecules, oxygen, water) which can passivate the defects, however this modification is temporary.[21] On the other hand, since oxygen is isovalent to sulphur, chemisorption of oxygen (during synthesis process) is expected to passivate sulphur vacancies without significantly modifying the crystal structure, and thus improve optoelectronic quality of the sample.[15,16,21,22] Also, the increased biaxial tensile strain in CVD samples due to the high synthesis temperature could induce a peak shift, and increase or decrease the PL intensity, depending upon the type and amount of strain present.[10,11,23]

To decouple the effect of defects, physisorption (and chemisorption), and strain on the optical properties of 2D materials, it is important to measure spectral signatures of defects directly. At RT, the luminescence due to defect bound excitons and adsorbates is unobservable, due to the thermalization of defects. At low temperatures (4 K, LT), non-radiative mechanisms due to phonon and carrier scattering are reduced, and defect PL can be observed as a broad low-energy peak (L-peak).[24] We also note the significant high-energy peak of S1 (> 1.9 eV), compared to S3 and S4, indicating that spin-split B-exciton is enhanced in S1. This may relate to higher density of defects in S1.[25]

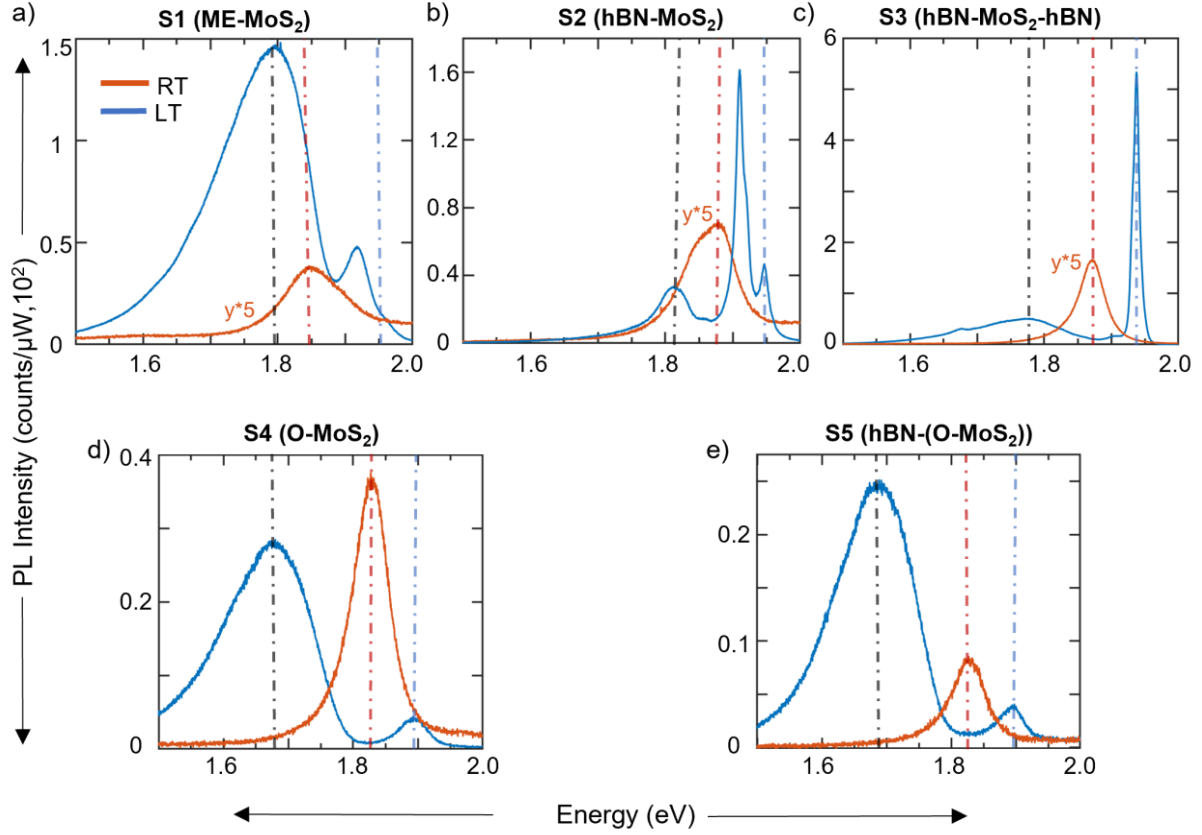

**Figure 2.** Room temperature (RT) and low temperature (LT, 4 K) PL spectra of samples a) S1 b) S2 c) S3 d) S4 e) S5. The red line indicates the RT spectra, while the blue line indicates the LT spectra in all samples. The intensity values of RT data in samples S1-S3 are multiplied by 5 for better visibility. The red, blue, and black dotted lines indicate the spectral positions of RT A-exciton, LT A-exciton and L-peak, respectively. The LT data in (a and c) have been reproduced from Ref [4]. The LT PL spectra of all samples were taken at 50 μW laser power.

For all samples at LT, we observe the L-peak, along with the delocalized A-exciton (X) and trion ($X^-$) peaks (Figure 2). Further, total PL intensity from the samples increases as the temperature is lowered, due to reduced phonon interactions (reduced scattering out of light cone) and non-radiative recombination (carrier scattering).[26] The blue-shift in peak position with temperature is due to increase in bandgap, as observed for most semiconductors.[27–29] For all samples, the linewidth of A-exciton reduces with temperature due to reduced phonon-induced homogeneous broadening. Further, hBN covering (S2) and encapsulation (S3) also improves sample quality, as evidenced by narrower linewidths (c.f. S1), but still not approaching the homogenous linewidth (~ 2 meV).[30–32] Interestingly, there is no evident shift of peak position and exciton peak intensity in S5 at LT, as compared to S4. This suggests that unlike the ME sample, the properties of O-CVD samples are stable, suggesting oxygen is chemisorbed in the O-CVD samples. Moreover, the L-peak PL intensity becomes narrower in S5, which could be due to reduced hydrocarbon contamination after covering. The linewidths and peak positions for all samples are summarized in Table 2 (also see Supplementary Section XIV). Anomalously, at LT, A-exciton PL is stronger in S1 compared to S4, whereas the reverse trend is observed for RT. Also, A-exciton peak in S4 is shifted by ~ 60 meV (compared to S1-S3). We will discuss this shift in detail in the later sections of the manuscript.

| Sample | Peak position for A-exciton (A-trion), meV | Peak position for L-peak | FWHM (A-exciton) (meV) |
|---|---|---|---|
| S1 | 1948 (1916) | 1797 | 50 |
| S2 | 1945 (1912) | 1818 | 20 |
| S3 | 1939 (1904) | 1766 | 15.6 |
| S4 | 1891 | 1672 | 60 |
| S5 | 1890 | 1679 | 62 |

**Table 2**. Extracted values of L-peak and A-peak from low temperature (LT) PL spectra of samples S1-S5.

The L-peak has previously been attributed to various mechanisms, including defect-bound excitons (single and bi-sulphur vacancies), adsorbates, and vacancy charge-transfer excitonic complexes with hydrocarbons.[7,33] The L-peak in all samples is a broad peak, but is visibly asymmetric, and is most likely comprised of two or more broad spectral peaks. For samples S1-S3, the position of the L-peak is at 1.79 $\pm$ 0.03 eV, with the shifts discussed later in terms of modified defects. The low-energy tail of the L-peak also exists in all samples, with varying intensity. The lower intensity of L-peak in S2 and S3, compared to S1, is attributed to lower extent of vacancy-hydrocarbon complexes and charged O adatoms in S2 and S3, as discussed later.

The O-CVD samples (S4 and S5) are drastically different compared to S1-S3. Firstly, there is a large shift in the L-peak position in S4 and S5 by ~ 130 meV (compared to S1-S3) at LT. The L-peak was observed to be slightly narrower for S4, compared to S1. The drastic shift in L-peak position is attributed to two effects. First is the lack of higher-energy defects due to passivation by oxygen chemisorption. The second is the tensile strain present in CVD samples due to high temperature growth. It is important to note that strain alone may not shift the L-peak, as seen for the smaller shift of ~ 60 meV for A-exciton between S1 and S4. Thus, both strain and defects are relevant. Even though both S1 and S4 have finite density of defects, the nature of defects can be different. In the next part of the manuscript, we understand the combined effect of strain, defects, and synthesis conditions on the optical properties of ML $MoS_2$.

| Defect | Formation energy (eV) | | | |
|---|---|---|---|---|
| | Our work | | Literature values | |
| | Mo rich | S rich | Mo rich | S rich |
| $V_S$ | 1.53 | 2.90 | $1.56^{34}$ | $2.93^{34}$ |
| $V_{Mo}$ | 7.52 | 4.77 | $7.27^{34}$ | $4.79^{34}$ |
| $V_{S+S}$ | 3.02 | 5.76 | $2.38^{12}$ | $5.15^{12}$ |
| $V_{Mo+S+S+S}$ | 6.06 | 7.43 | $6.29^{34}$ | $7.93^{34}$ |
| $O_S$ | -2.80 | -1.40 | $-2.71^{12}$ | $-1.88^{35}$ |
| $O_{ad}$ | -0.65 | -0.65 | $-0.81^{35}$ | $-0.81^{35}$ |
| $O_S + O_{ad}$ | -3.43 | -2.03 | - | - |
| $O_{ad}$ (q=-2) | -1.63 to 1.82 | -1.63 to 1.82 | - | - |
| $O_{ad}$ (q=+2) | -1.98 to 1.46 | -1.98 to 1.46 | - | - |

**Table 3**. Comparison of calculated formation energy with reported values of different defects in monolayer $MoS_2$ in Mo-rich and S-rich conditions at 0 K. For the charged $O_{ad}$ defects, the range of values reported represent the range of formation energies as the Fermi energy varies from the valence band edge to the conduction band edge.

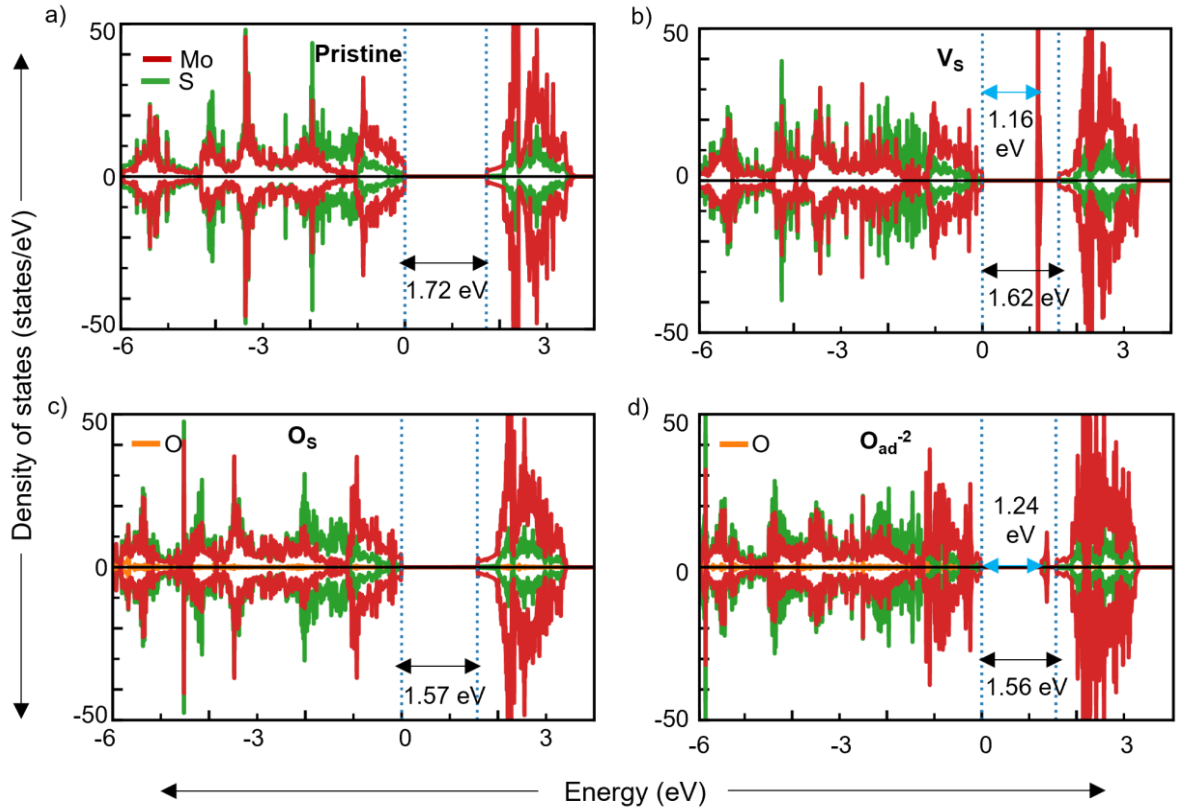

**Figure 3.** Calculated electronic density of states for a) pristine monolayer (ML)-MoS$_2$, and ML-MoS$_2$ with b) V$_S$, c) O$_S$ and d) O$_{ad}$ (q = -2). Bandgap magnitudes (black arrows) are indicated, along with defect levels (blue arrows) for V$_S$ and O$_{ad}$ (q = -2).

To understand the nature of defects contributing to the broad L-peak, electronic density of states (DOS) calculations were carried out, as shown in Figure 3 and Supplementary Figure S-VI. In all the DOS plots, the red, green, and orange lines indicate the Mo-d, S-p and O-p state respectively. The dotted blue vertical lines mark the band edges, and the band gap magnitudes are denoted by the text within the panels. Note that O$_{ad}$ and O$_S$ represent an oxygen atom adsorbed on top of a S atom and an O anti-site formed in a vacant S site, respectively. We have also compiled the formation energies of several defects in Table 3, and compared our calculated values with available literature values. We also predict the formation energies of O$_S$ + O$_{ad}$ and O$_{ad}$ (q = +2, -2), for which literature values were not found.

The calculated bandgap of pristine ML-MoS$_2$ is 1.72 eV, as shown in Figure 3a, which is consistent with previous calculations.[34] The DOS of V$_S$ confirms the presence of in-gap defect states (1.16 eV from the valence band edge in Figure 3b). This defect state can be passivated by O$_S$ (Figure 3c), as well as $O_S + O_{ad}^{top}$ (Supplementary Figure S-VI b) wherein the $O_2$ molecule dissociates at a S vacancy (see Supplementary Figure S-II d; 'top' superscript indicates the location of the adsorbed O), in agreement with Ref [21]. Neutral $O_{ad}$ in ML-MoS$_2$ does not show any defect states within the band gap, however its charged counterparts, q = +2 and –2, which are also stable within the MoS$_2$ bandgap (see Supplementary Figure S-VI d and Figure 3d), show the presence of shallow defect states. The defect states in $V_S$ and charged (q = -2 and +2) $O_{ad}$ are dominated by the Mo d-states.

The combined effect of biaxial strain and defects is illustrated using band structure plots in Figure 4, to visualize deviations from the direct band gap nature of the pristine unstrained structure. The change in band gap and defect energy levels in pristine MoS$_2$ as well as defect-containing MoS$_2$ with $V_S$, and $O_{ad}$ (q=-2) (i.e., those defects that have low formation energy and exhibit an in-gap defect state) at 0.5% and 1% strain conditions are included in Figure 4a. The band structures of MoS$_2$ with a few other defects and applied strain are compiled in Supplementary Figure S-VII. The dotted black lines in each plot represents the band edges. The DOS of structures deformed by biaxial strain, which are consistent with our band structure calculations, are compiled in Supplementary Figure S-VIII. We find that the unstrained pristine ML-MoS$_2$ possesses a direct bandgap of 1.72 eV at the K point, in agreement with previous studies.[36] In the case of pristine structure deformed by biaxial tensile strain of 0.5%, the bandgap is no longer direct (1.61 eV), as the valence band maximum (VBM) is now at the Γ point (K-K direct gap at 1.70 eV). Similarly, for the 1% strained structure, the valence band maximum does not lie at the K point, with a decrease in band gap to 1.49 eV (K-K direct gap at 1.66 eV).

In the band structure plots of $V_S$, two closely degenerate in-gap states (at 1.17 eV from the VBM) are found, with the VBM at Γ point, which is in line with Ref[36]. With increasing strain of 0.5% and 1%, the location of the defect state decreases to 1.11 eV and 1.05 eV from the VBM, respectively. Secondly for the case of $V_S$, the gap also becomes indirect with increasing strain. We find that with the $O_S$ in the unstrained case, the direct band gap of ML MoS$_2$ is not preserved, and with increasing strain %, the band gap of $O_S$ decreases (Supplementary Figure S-VII). Further, in the unstrained neutral $O_{ad}$ case, a direct band gap of 1.74 eV (similar to that of the pristine structure) is found (Supplementary Figure S-VII), which decreases and becomes indirect with strain. On the other hand, the negatively charged (q =

-2) $O_{ad}$ (Figure 4a) shows an in-gap defect state near the CBM, which gets closer to the CBM with increasing strain percentage. For the negatively charged $O_{ad}$, the band gap remains direct, in contrast with other defects.

A consolidated bar chart showing the changes in band gap and defect states for each system is displayed in Figure 4b. In the histogram, we indicate the direct band gap magnitudes, since PL probes the direct transitions and ensuing luminescence. We also indicate the indirect band gap (using dotted lines) for completion. The decrease in direct band gap between the unstrained and 0.5% strained pristine ML-$MoS_2$ is ∼ 20 meV, and that between the unstrained and 1% strained pristine ML $MoS_2$ is ∼ 60 meV. In the case of $O_S$, the bandgap reduces to 1.46 eV and 1.34 eV at 0.5% and 1% strain respectively. For the neutral $O_{ad}$, the band gap reduces to 1.62 eV and 1.50 eV at 0.5% and 1% strain respectively. For charged $O_{ad}$ (q=-2), the bandgap reduces to 1.56 eV and 1.54 eV for 0% and 0.5% strain respectively, and remains the same for 1% strain.

For defect levels, optical transitions do not have to be vertical in energy-momentum space, and thus we plot defect energy separation from the VBM (or CBM). The difference in energy of $V_S$ defect state with respect to the VBM is 50 meV and 120 meV for the 0.5% and 1% strain, respectively, in comparison to the unstrained $V_S$ defect. We suggest that the changes in the band gap and the defect states within the gap contribute to the shifts observed in the exciton peaks in our measured PL.

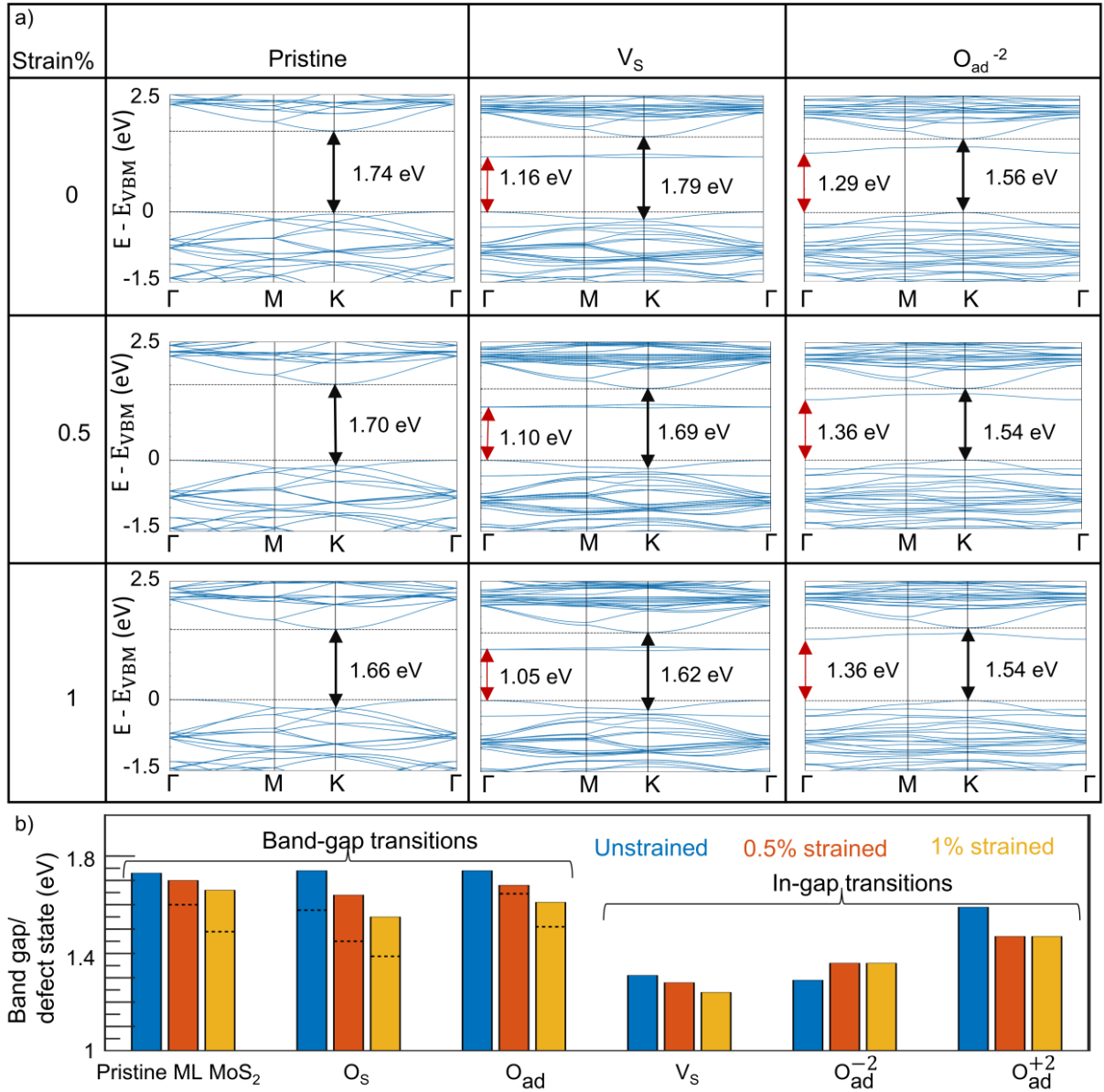

**Figure 4.** a) Band structures for Pristine ML-MoS$_2$, with V$_S$ and O$_{ad}$ (q = -2), under 0%, 0.5%, and 1% applied biaxial tensile strain. Black arrows indicate optically relevant direct bandgap transitions (K-K) and red arrows indicate defect-induced in-gap transitions. b) Bar chart representing the variation of bandgap and the positions of defect-induced in-gap states with different defects and applied strain. For bandgaps, solid horizontal lines indicate direct bandgap (K-K) and dotted lines indicate indirect bandgap (K-Γ) transitions.

Let us now focus on understanding the origin of L-peak, given insights from our DFT calculations. Excitation laser power induced saturation of PL intensity can be observed if luminescence is contributed by defect-bound excitons, while the free (delocalized) excitonic peak intensity will increase linearly with power. Thus, PL was performed as a function of excitation laser power. In Figure 5a, 5b, we show data for S1 and S4 respectively (see Supplementary Section VII for all samples). We observed that the A-exciton PL intensity scales nearly linearly with power for all samples. On the other hand, L-peak shows saturation with power. Specifically, log of integrated intensity v/s log of laser power shows sub-linear (slope < 1) dependence for L-peak, confirming our assignment to defect peak (see inset of Figure 5a, b). For different samples, the slope varies, as summarized in Table 4. Further, the L-peak saturates fastest for S3, indicating low defect densities, since the peak saturation depends on available states for radiative recombination. Interestingly, L-peak saturates faster for S4, c.f. S1. The increase in RT PL of S4 and S5 c.f. S1, and the faster saturation of L-peak in LT PL, indicate that S4 and S5 have lower density of defects compared to S1. Changes in the peak behaviors indicate that defect density and effect of adsorbates vary with the sample preparation techniques.

| Sample Label | L-peak coefficient | A-peak coefficient |
| --- | --- | --- |
| S1 | 0.84 | 1.08 |
| S2 | 0.94 | 0.93 |
| S3 | 0.83 | 0.99 |
| S4 | 0.78 | 0.9 |
| S5 | 0.74 | 1.09 |

**Table 4**. Coefficients of L-peak and A peak from log-log plot of integrated intensity v/s excitation power for different samples.

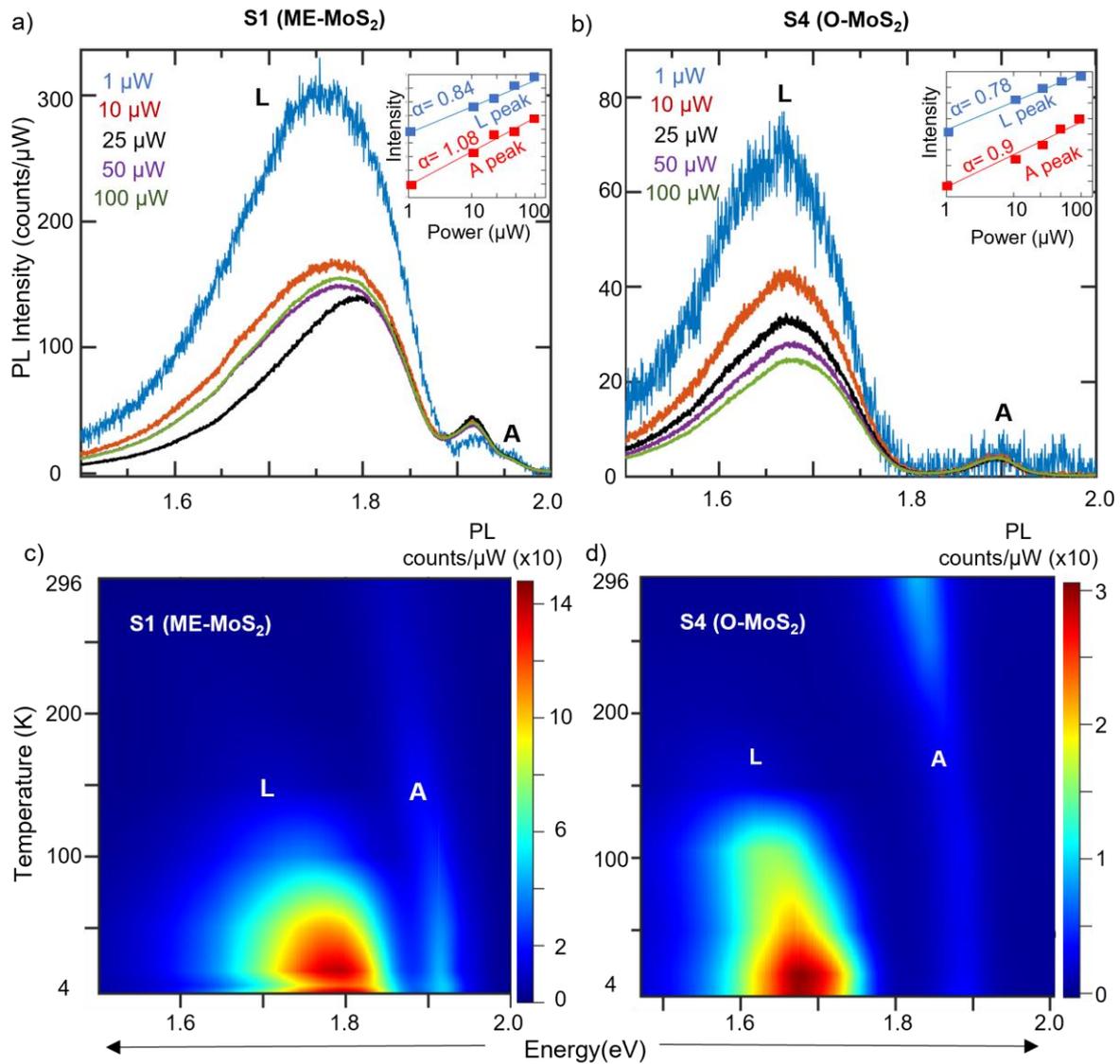

**Figure 5.** Power dependent PL spectra of a) S1 and b) S4 at 4K. The PL spectra is normalized with power to illustrate saturation behavior of L-peak. Inset shows log-log plot of intensity v/s excitation power, power coefficients are also mentioned. Surface plots showing evolution of PL spectra with temperature for c) S1 and d) S4. The L-peak emerges below 150 K.

Temperature dependent PL spectroscopy shows that L-peak is observable below 150 K for both S1 and S4 (Figure 5 c, d). For L-peak, increasing PL intensity with decreasing temperature is indicative of defect potential-trapped bound excitons (Figure 5c, d). With decreasing temperatures, an increase of A-exciton PL intensity is observed for all ME samples (Figure 5c, and Supplementary Figure S-X a, b). Remarkably, an anomalous decrease in A-exciton PL intensity is observed for S4 and S5 (Figure 5d, and Supplementary Figure S-X c), attributed to only defects in earlier work.[37] From our band structure calculations, we observed that strain can change the nature of bandgap, transitioning from direct to slightly indirect (Figure 4). Shift of the valence band maximum towards the Γ point with increasing strain, and the possibility of MoS$_2$ ML to become indirect (for strain > 1%), has been previously reported.[38] Further, defects can also cause a change in bandgap (Figure 4). The decrease in the A-exciton PL in S4 and S5 with decreasing temperature is thus attributed to slightly indirect nature of strained samples, and with contributions from defects.

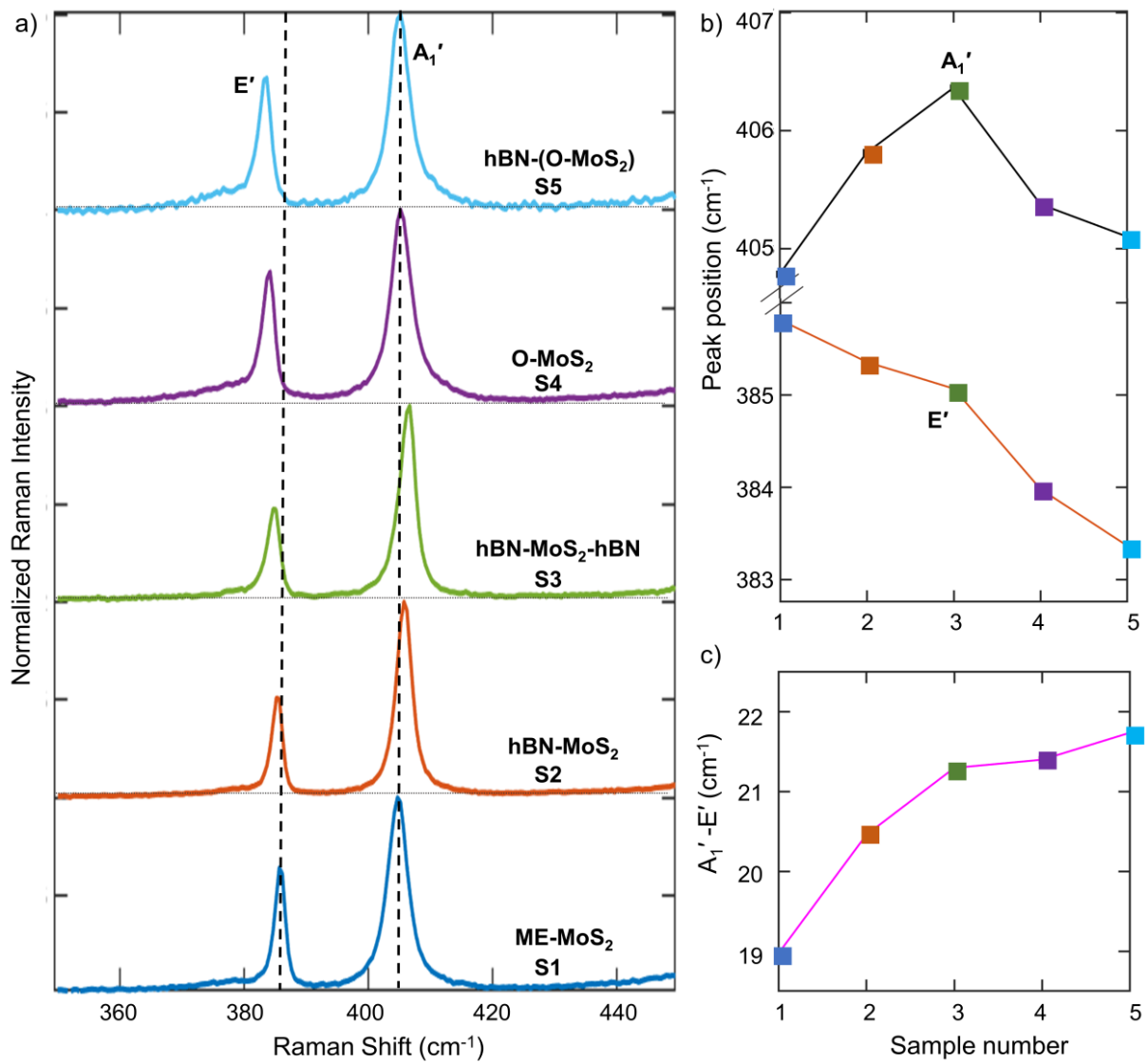

**Figure 6.** a) Comparison of Raman spectra for samples S1-S5. The Raman intensity values of all samples are background subtracted, and then normalized to their respective A$_1'$ peak intensity. Raman data without normalization can be found in Supplementary Figure S-XI. The dotted lines indicate the peaks for S1. b) Peak

positions of $A_1'$ and $E'$ vibrational modes in Raman spectra of samples S1-S5 c) Difference between $A_1'$ and $E'$ peaks in Raman spectra of samples S1-S5.

To further decouple strain, doping and defect density in the samples, we performed Raman spectroscopy (at RT). As seen in Figure 6a, in-plane $E'$ peak is increasingly blue shifted for samples from S1 to S5, indicative of increasing strain in the sample for different processing conditions. Interestingly, out-of-plane $A_1'$ peak shifts progressively with hBN covering and hBN encapsulation in S2 and S3 respectively, which originates from modified vdW interaction and screening, as well as reduced doping from substrate and environment.[9,39] We emphasize again that S3 is a nearly pristine sample (low doping and defects), and may be considered as a good reference for comparison. On the other hand, $A_1'$ peak of S4 and S5 remains nearly same as S1, thus ruling out the presence of increased doping in the CVD sample (c.f. ME). The large shift in $E'$ for S4 and S5 (c.f. S1) is attributed to thermal strain developed in the sample during high temperature growth. Considering the peak shift between S1 and S4 to be ~ 2 cm$^{-1}$, and the strain induced $E'$ peak shift ~ 4.2 cm$^{-1}$ per 1 % biaxial strain (biaxial)[11,40,41], the strain is estimated to be ~ 0.5%, which is large. Thus, the peak shift is attributed to strain and defects-induced distortion of lattice induced during O-CVD. Further, a shoulder peak is observed for all samples around 379 cm$^{-1}$, which is referred to as LO peak (ref), and is indicative of density of defects.[42] The peak difference between $E'$ and $A_1'$ increases from S1-S5, and as discussed, is indicative of strain and changes in the substrate-sample interaction. Thus, the modifications to properties and quality of sample due to synthesis procedure and post-processing can be clearly understood.

We then perform XPS on samples S1 and S4 to measure the difference in nature of oxygen bonding. We confirm the presence of chemisorbed oxygen in O-CVD sample by measuring high percentage of Mo(VI 3d$_{3/2}$)-O bonds in S4 (see Supplementary Section XI for fitted XPS spectra and analysis).

The differences in the optical signatures of the L-peak in different samples suggest different combinations of defects and strain in the samples. With combined knowledge from DFT band structure calculations and LT PL spectra, we attribute the L-peak in S1 to a combination of $V_S$, $O_{ad}$ (q=±2) and hydrocarbon complexes. Reduction in the L-peak intensity after hBN covering and encapsulation is attributed to reduction of charged O adatoms and hydrocarbon complexes (due to transfer procedure and hBN covering). Thus, in S2 and S3, the L-peak would be primarily contributed by $V_S$ with minor contributions from hydrocarbon complexes and charged O adatoms. Further, the nature of defects in O-CVD samples is very different from ME samples. For example, the formation energy of O adatom becomes positive at 1023 K (i.e. growth temperature), as shown in supplementary Figure S-IV. Interestingly $O_S$, which does not contribute in-gap states, has a lower formation energy at 1023 K compared to charged O adatoms. Thus, the L-peak in O-CVD samples originates from $V_S$ and hydrocarbon complexes. This is consistent with further reduction in linewidth upon hBN covering (see Supplementary Figure S-XIV). The role of charged complexes is also supported by gate-dependent measurements performed earlier by Chen et al.[6] The nature of defects and effect on optical properties are summarized via a schematic in Supplementary Figure S-XV.

**Conclusion:**

In conclusion, we uncover the physical origins of L-peak by measuring a comprehensive set of samples designed to decouple defects, strain, and dielectric environment. DFT calculations of defect formation energies and band-structure are performed to understand the nature of defects (and strain) and effect on optical properties. Anomalously high A-exciton PL of O-CVD samples (c.f. ME) at RT, but reduced PL at LT, is attributed to low density of defects and indirect gap transition due to synthesis-induced strain. Drastic shift of ~130 meV for L-peak in O-CVD samples (c.f. ME) is attributed to a combination of tensile strain and absence of charged oxygen adatoms. Comparing bandgap values obtained through DFT with the peak shifts observed in PL, we uncover that L-peak in ME samples originates from a combination of sulphur vacancies, charged O adatoms and hydrocarbon complexes. For O-CVD samples, L-peak originates only from sulphur vacancies and hydrocarbon complexes. The conclusions are well supported by Raman measurements, power-dependent PL, and temperature dependent PL. Presence of chemisorbed oxygen in O-CVD samples is confirmed by XPS.

Importantly, the role of hBN encapsulation in improving optical quality is clarified, and attributed to reduction in charged O adatoms and hydrocarbon complexes. This helpful effect of encapsulation holds for both ME and O-CVD samples. Finally, we emphasize that strain and defects should be considered together for their effect on optoelectronic properties. Thus, careful control and choice of synthesis and post-processing conditions is the key in obtaining materials with desired optical properties for optoelectronics and quantum technologies.

**Methods:**

**Sample preparation:** ML MoS$_2$ sample.

Oxygen assisted chemical vapor deposition (O-CVD)

O-CVD was done using sulphur and MoO$_3$ powder precursors kept in the first and second heating zones of a three-zone furnace at 200°C and 530°C respectively. 285 nm prime SiO$_2$/Si substrate was kept vertically in the third zone at 750°C to ensure uniform precursor concentration along the substrate. The tube was ramped up to the respective temperatures in 30 minutes and maintained for 20 minutes for growth. The sulphur boat was kept outside the first zone during heating and later pushed in using magnets right after the temperatures were attained. 100 sccm N$_2$ was used as carrier gas with 2 sccm of oxygen to prevent sulphurization of MoO$_3$ in the precursor boat. Oxygen flow was stopped 5 minutes after the set temperatures were reached to prevent etching of the as-grown sample and excessive doping. The furnace was opened after 20 minutes of growth and cooled to room temperature in the presence of 200 sccm N$_2$.

Mechanical Exfoliation and heterostructure preparation

MoS$_2$ (2D semiconductors) and hBN flakes (NIMS, Japan) were prepared by micro-mechanical exfoliation of respective bulk crystals using scotch tape method. Monolayers of MoS$_2$ were identified using optical contrast method. We used PDMS-PPC based transfer method to prepare hBN covered and hBN encapsulated ML MoS$_2$ samples.[4] After the heterostructure is prepared, the sample was annealed in Nitrogen atmosphere in glovebox for 3 hours at 250°C to reduce organic contaminants and improve the heterostructure interface. hBN covered CVD samples were prepared by all dry viscoelastic stamping method.[43] hBN flakes are exfoliated onto PDMS sheet and transferred onto CVD flakes at room temperature. The heterostructure was annealed at 150°C for 10 minutes in glovebox to improve coupling between layers and reduce organic contaminants.

**DFT calculations**

The electronic ground states of pristine ML-MoS$_2$ and its defective configurations were calculated with DFT, as implemented in the Vienna ab initio simulation package (VASP)[44,45] and employing the projector-augmented-wave (PAW)[46] potentials for describing the core electrons. We expanded the plane-wave basis set up to a kinetic energy cut-off of 520 eV and utilized the strongly constrained and appropriately normed (SCAN)[47,48] functional to describe the electronic exchange and correlation (XC). We sampled the irreducible Brillouin zone on a well converged Γ-centered $k$-point mesh with a density of 48 $k$-points per Å (e.g., a 4 Å lattice parameter will be sampled using 12 $k$-points in the corresponding reciprocal space direction) and integrated the Fermi surface with Gaussian smearing with a width of 0.05 eV. For both pristine and defective ML configurations, we allowed only the ionic positions to relax till the total energies and atomic forces converged below $10^{-5}$ eV and $|0.01|$ eV/Å. We used a 4 x 4 x 1 supercell for all defect calculations upon verifying the convergence of DFT defect formation energies (within ~0.1 eV) for the neutral sulfur vacancy ($V_S$) defect (Supplementary Figure S-III). The distance between two periodic images in the out-of-plane direction is 21 Å, for both the pristine unit cell and defective supercell MoS$_2$ configurations.

We used the inorganic crystal structure database[49] to obtain the initial configuration of Mo and S atoms in pristine ML-MoS$_2$. The strained structures for pristine ML-MoS$_2$ and defective ones were generated from their corresponding unrelaxed structures. We applied biaxial strains (i.e., along the *a-b* plane), with magnitudes of 0.5% and 1% compared to the original lattice parameters, using the pymatgen package.[50,51] Note, only the ionic positions were relaxed for all the strained structures.

For calculating the electronic density of states (DOS), we performed a single self-consistent-field (SCF) calculation, on the relaxed lattice geometry, for the cases of pristine, strained, and defective ML-MoS$_2$, with a $k$-mesh density of 144 $k$-points per Å (i.e., 3 × the density used in structure relaxations). Note that we used the tetrahedron smearing scheme[52] for calculating all electronic DOS. To calculate the dielectric constant of pristine ML-MoS$_2$ with the SCAN functional, we introduced small symmetrically-distinct perturbations of 0.015 Å to the SCAN-relaxed atomic positions using the finite displacement method to capture the ionic relaxation contributions to the dielectric tensor. Also, we calculated the ion-clamped static dielectric tensor via the self-consistent response to a finite electric field, equivalent to a magnitude of 0.01 eV/Å, in all three directions. The electronic band structures in pristine and defective ML-MoS$_2$ were calculated with SCAN along the well-known –M–K– path in the reciprocal space.[51] Note that we used the Latimer Munro scheme[53] to generate a list of high symmetry *k*-points for the band structure calculation, from which the Γ, M and K points were selected. We used the pymatgen[50,51] package for pre- and post-processing our DFT calculations.

The formation energy for any defect is given by,

$$E_{defect}^f = E_{defect} - E_{pristine} - \sum_i n_i \mu_i + qE_F + E_{corr}$$

where $E_{defect}$ and $E_{pristine}$ are the total SCAN-calculated energies of defective and pristine ML-MoS$_2$ respectively. $n_i$ is the number of atoms of species being added (> 0) or removed (< 0), while $\mu_i$ represents the corresponding chemical potential. $E_F$ and $E_{corr}$ are the Fermi energy of pristine ML-MoS$_2$ and the electrostatic correction, respectively, which are appropriate for defects with non-zero charge. As ML-MoS$_2$ is anisotropic, we used the scheme proposed by Kumagai and Oba,[54] as implemented in the python charged defect toolkit (PyCDT) to account for $E_{corr}$.[55] A representative calculation of $E_{corr}$ term, for the case of charged V$_S$, is given in the Supplementary Section IV.

**Raman and PL measurements:** Raman measurements were carried out in a HORIBA LabRamHR Raman set up using 532 nm laser, 1800 grating lines/mm and 100x objective. The laser power used for Raman measurements was $\leq 100 \,\mu W$. Room temperature PL mapping was done using Witec Alpha 300 system using 532 nm laser, 100x objective and 600 grating lines/mm. All other PL measurements were done using a customized set up consisting of Montana cryo-system, Andor spectrometer (300 grating lines/mm) and silicon CCD with 482 nm excitation laser, 50X objective and laser power $\leq 100 \,\mu W$.

**XPS measurements**: XPS measurements were carried out using Thermofisher K-α with a 1.4 keV X-ray source, 120 μm probe and 50eV pass energy. The analysis was done using CASA software.

ASSOCIATED CONTENT

The following files are available free of charge.

Supplementary Information (PDF)

AUTHOR INFORMATION:

**Corresponding Author:**

*Akshay Singh, aksy@iisc.ac.in

**Author Contributions**

KSK, AKD, MV and AS developed the experimental framework. HSH and GSG performed the DFT of defects. KSK and AKD performed the optical experiments, with assistance from MV. KSK and MV performed the O-CVD synthesis of ML $MoS_2$. KSK and VK performed the XPS measurements. KSK performed the data analysis, with assistance in PL analysis by MV and AKD. KW and TT provided the hBN bulk crystals. KSK and AS discussed and prepared the manuscript, with contributions from all authors.

**Data Availability**

All data is available upon reasonable request.

ACKNOWLEDGMENTS

AS would like to acknowledge funding from Indian Institute of Science start-up and SERB grant (SRG-2020-000133). AKD would like to acknowledge Prime Minister's Research Fellowship (PMRF). KSK would like to acknowledge DST-INSPIRE Fellowship. The authors also acknowledge Micro Nano Characterization Facility (MNCF), Centre for Nano Science and Engineering (CeNSE) and XPS facility, Department of Inorganic and Physical Chemistry (IPC), IISc for use of characterization facilities. T.T. acknowledges support from the JSPS KAKENHI (Grant Numbers 19H05790 and 20H00354) and A3 Foresight by JSPS. GSG acknowledges the computational resources provided by the Supercomputer Education and Research Centre (SERC), IISc. A portion of the calculations in this work used computational resources of the supercomputer Fugaku provided by RIKEN through the HPCI System Research Project (Project ID hp220393). We acknowledge National Supercomputing Mission (NSM) for providing computing resources of 'PARAM Siddhi-AI', under National PARAM Supercomputing Facility (NPSF), C-DAC, Pune and supported by the Ministry of Electronics and Information Technology (MeitY) and Department of Science and Technology (DST), Government of India.

ABBREVIATIONS

2D, two dimensional; TMDs, transition metal dichalcogenides; ML, monolayer; RT, Room Temperature; LT, Low Temperature (4 K); PL, photoluminescence; DFT, density functional theory; XPS, X-ray photoelectron spectroscopy

**Supporting Information**

**Towards a comprehensive understanding of the low energy luminescence peak in 2D materials**

Keerthana S Kumar[1], Ajit Kumar Dash[1], Hasna Sabreen H[2], Manvi Verma[1], Vivek Kumar[1], Kenji Watanabe[3], Takashi Taniguchi[4], Gopalakrishnan Sai Gautam[2], Akshay Singh[1, *]

[1]Department of Physics, Indian Institute of Science, Bengaluru, Karnataka -560012, India

[2]Department of Materials Engineering, Indian Institute of Science, Bengaluru, Karnataka -560012, India

[3]Research Center for Functional Materials, National Institute for Materials Science, Ibaraki 305-0044, Japan

[4]International Center for Materials Nanoarchitectonics, National Institute for Materials Science, Ibaraki 305-0044, Japan

*Corresponding author: aksy@iisc.ac.in

**I) Optical microscope images of exfoliated samples and hBN covered oxygen assisted chemical vapor deposition (O-CVD) sample:**

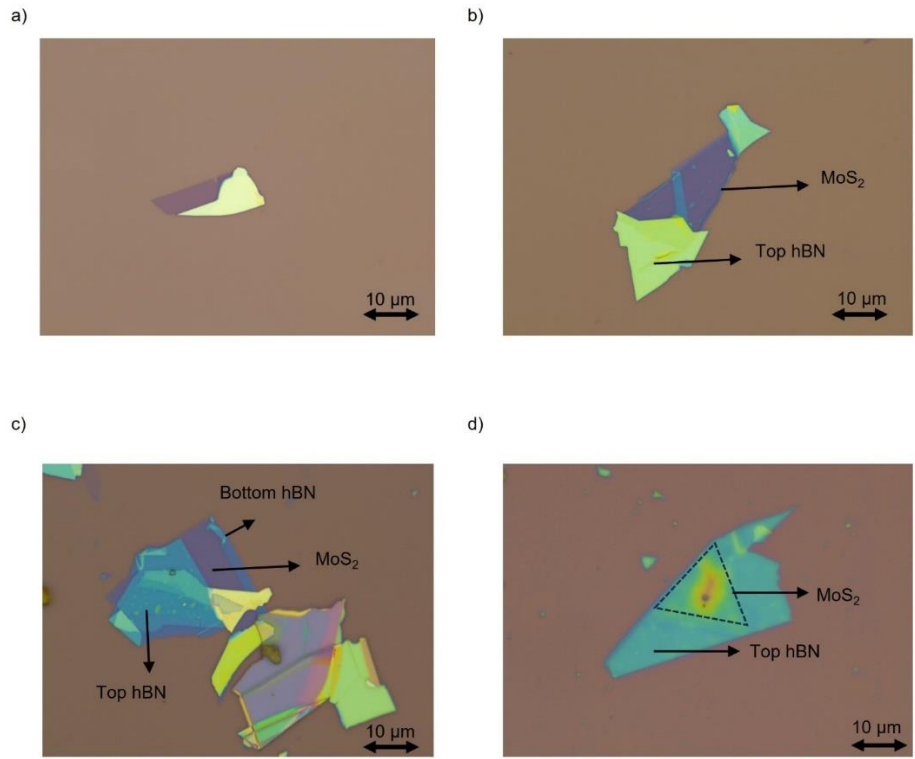

**Figure S-I**. Optical microscope images of monolayer (ML)-$MoS_2$ samples a) S1 b) S2 c) S3 d) S5 discussed in Main text. The light purple contrast corresponds to monolayer $MoS_2$ as labelled in the images, and is clearly distinguishable from the substrate contrast as well as the contrast of bulk $MoS_2$ and hBN.

**II) Schematics of different types of defects in monolayer (ML)-MoS$_2$:**

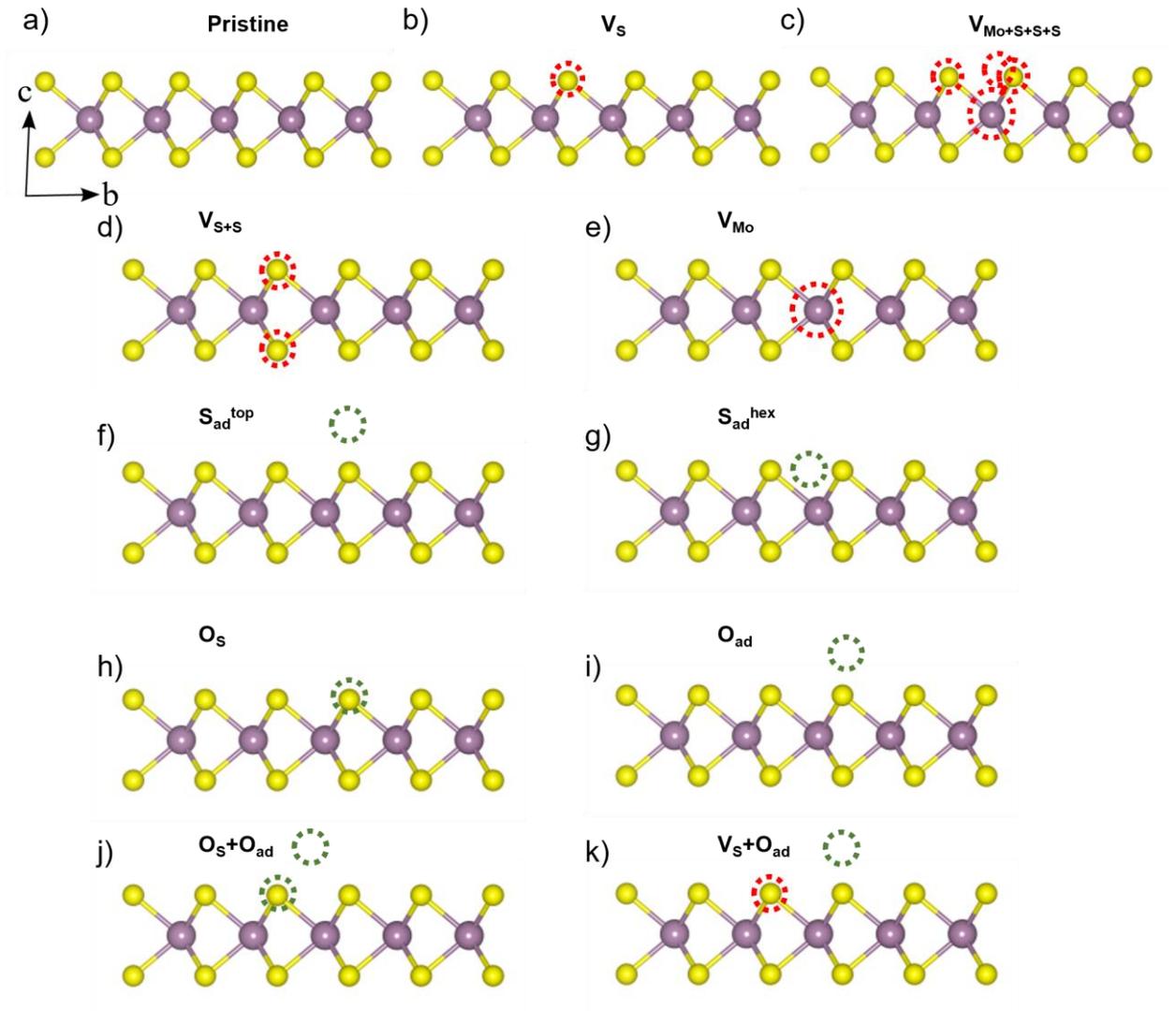

**Figure S-II.** Initial atomic structures of (a) pristine ML-MoS$_2$, ML-MoS$_2$ with (b) mono S vacancy (V$_S$), (c) Mo-S$_3$ vacancy cluster (V$_{Mo+S+S+S}$). (d) di-S vacancy (along the top and bottom of the ML–V$_{S+S}$), (e) mono Mo vacancy (V$_{Mo}$). Red dotted circles indicate that particular atom is missing, (f) S adatom on top of S (S$_{ad}^{top}$), (g) S adatom in the hexagonal void between S atoms (S$_{ad}^{hex}$), (h) O antisite at S (O$_S$), (i) O adatom on top of S (O$_{ad}$) , (j) O antisite at S and O adatom on top of S (O$_S$ + O$_{ad}$) and k) S vacancy and O adatom on top of S (V$_S$ + O$_{ad}$). The yellow and purple spheres indicate S, Mo respectively. For each structure, cross-sectional ($b - c$ plane) is shown. The adatoms are initialized at a distance of 2 Å from the nearest atom on the ML along the $c$-axis.

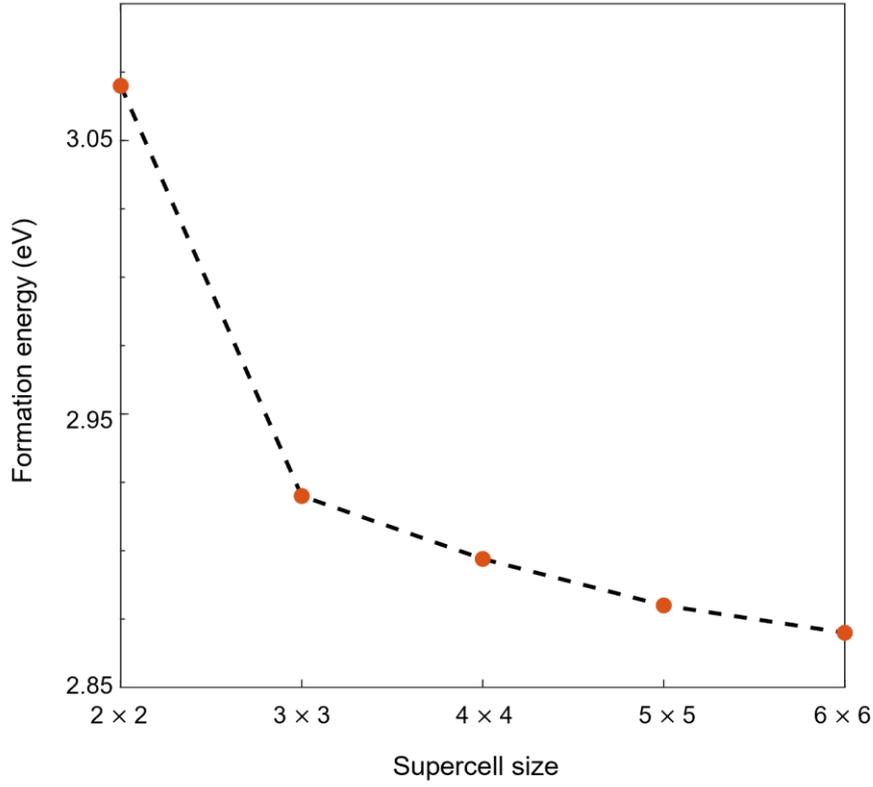

**Figure S-III.** Change in formation energy with the supercell size for a sulfur vacancy ($V_S$) defect in monolayer (ML) $MoS_2$. The formation energy converges to within ~0.1 eV at the 4×4 supercell, with respect to the larger 6×6 cell, which is the typical convergence criteria for defect formation energies using computations.[1,2] All subsequent calculations are hence carried out in a 4×4 cell.

### III) Formation energy of defects in monolayer $MoS_2$ in different growth conditions:

The defect formation energies were calculated using

$$E_{defect}^f = E_{defect} - E_{pristine} - \sum_i n_i \mu_i + qE_F + E_{corr}$$

Where $E_{defect}$ and $E_{pristine}$ are the total DFT-SCAN[3,4] derived energies of defective and a pristine supercell respectively, $n_i$ is the number of atoms of atoms of species i being added (> 0) or removed (< 0), $\mu_i$ represents the chemical potential of species i, and $E_F$ and $E_{corr}$ are the Fermi energy of the pristine supercell and electrostatic correction, respectively, which are appropriate for defects of non-zero charge

q. As ML-MoS$_2$ is anisotropic, we use the scheme proposed by Kumagai and Oba[5], as implemented in the Python charged defect toolkit (PyCDT) to account for the electrostatic corrections [6].

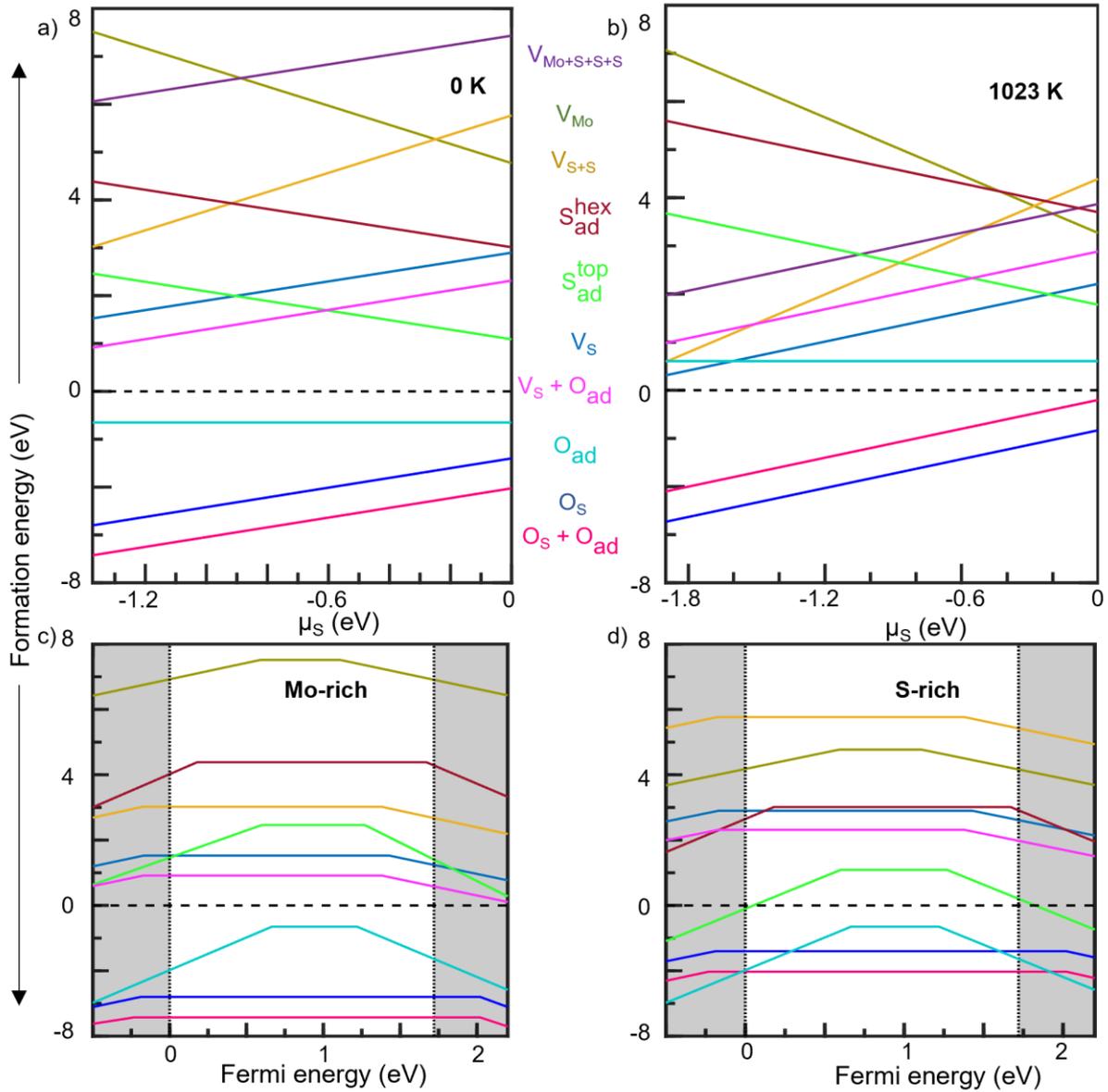

**Figure S-IV.** Formation energy of different neutral defects at a) 0 K and b)1023 K, as a function of the sulfur chemical potential ($\mu_S$). Defect transition levels for various charged defects at 0 K under c) Mo-rich ($\mu_S$ = -1.37 eV) and d) S-rich conditions ($\mu_S$ = 0 eV). The dotted black lines indicate the calculated band edges of pristine ML-MoS$_2$, which are separated by the bandgap (1.72 eV). The shaded regions indicate the valence and conduction bands of pristine ML-MoS$_2$.

The formation energy of defects at different temperatures can be estimated by determining the chemical potential ($\mu_i$) of ML-MoS$_2$, the constituent components Mo and S and the impurity component (O) at a temperature T as

$\mu_i (T, p) = \mu_i^o (T_o, p_o) + H (T, p_o) - TS (T, p_o) + RT \ln(p_i/p_0)$

$$\mu_{comp}^o = \frac{1}{2} G_{comp} = \frac{1}{2} H_{comp} - \frac{1}{2} TS_{comp}$$

Where $\mu_i^o (T_o, p_o)$ is the reference chemical potential of the compound (component) set to the 0 K DFT energy, $p_i$ is the partial pressure of the component and $p_o$ is the reference pressure (1 atm). The terms H (T) and S (T) corresponds to the enthalpy and entropy, respectively sourced from the National Institute of Standards and Technology (NIST) database[7]. H (T) is set to 0 at 298 K for pure Mo, S and O, while it is assigned the formation energy of MoS$_2$ at 298 K.

The change in defect formation energy with chemical potential of S ($\mu_S$) for all the neutral defects at 0 K and at 1023 K (i.e., the temperature maintained during the O-CVD synthesis of ML-MoS$_2$) is plotted in Figure S-IV (a) and (b) respectively. The range of chemical potential for S ($\mu_S$) varies from -1.37 eV to 0 eV at 0 K to -1.91 eV to 0 eV at 1023 K. Table S-I lists the defect formation energies of various defects in ML-MoS$_2$ in both Mo- and S-rich conditions at 0 K and 1023 K. From Figure S-IV a, the formation energies of Mo vacancy, MoS$_3$ vacancy cluster, di-S vacancy, and S adatom (hex) are so high that they will likely not be easily formed in both S- and Mo-rich conditions at 0 K. S vacancy in Mo-rich conditions (1.52 eV) and S adatom in S-rich condition (1.09 eV) at 0 K are the most stable intrinsic point defects with the lowest formation energies as indicated in Figure S-IV a.

The 0 K defect formation energies are negative for all the three O-related defects ($O_S$, $O_{ad}$ and $O_S + O_{ad}$) suggesting their spontaneous formation. However, at 1023 K, the formation energy of $O_{ad}$ is positive and may not spontaneously form. Comparing the 0 K and 1023 K formation energy plots (Figure S-IV a, b), the vacancy defects show a decrease and the adatom defects show an increase in their formation energies with increase in temperature. The formation energy of $V_S$ decreases to 0.31 eV at Mo-rich conditions at 1023 K whereas the formation of $S_{ad}^{top}$ becomes difficult at higher temperatures due to the increased formation energy (1.78 eV). As we also notice a decrease in the formation of $V_{S+S}$ in Mo-rich conditions at 1023 K, S vacancies are the most abundant intrinsic point defects at these high temperatures. The predominant occurrence of S vacancy due to its lower formation energy is reported in Ref[8]. It is interesting to note that the formation energy of $V_{Mo+S+S+S}$ significantly reduces from 6.06 eV at 0 K to 1.97 eV at 1023 K in Mo-rich conditions. This decrease in formation energy of $V_{Mo+S+S+S}$ cluster, compared to $V_{Mo}$, illustrates that S vacancies can easily form around a Mo vacancy, thus stabilizing the $V_{Mo+S+S+S}$ cluster at high temperatures.

| Defect | Formation energy at 0 K | | Formation energy at 1023 K | |
|---|---|---|---|---|
| | Mo-rich | S-rich | Mo-rich | S-rich |
| $V_S$ | 1.52 | 2.90 | 0.31 | 2.21 |
| $V_{Mo}$ | 7.52 | 4.77 | 7.07 | 3.27 |
| $V_{S+S}$ | 3.02 | 5.77 | 0.59 | 4.39 |
| $V_{Mo+S+S+S}$ | 6.06 | 7.43 | 1.97 | 3.87 |
| $S_{ad}^{top}$ | 2.46 | 1.09 | 3.68 | 1.78 |
| $S_{ad}^{hex}$ | 4.38 | 3.01 | 5.60 | 3.70 |
| $O_S$ | -2.80 | -1.40 | -2.73 | -0.83 |
| $O_{ad}$ | -0.65 | -0.65 | 0.60 | 0.60 |
| $O_{ad}$ (q = -2) | -1.63 to 1.82 | -1.63 to 1.82 | -0.37 to 0.61 | -0.37 to 0.61 |
| $O_{ad}$ (q = +2) | -1.98 to 1.46 | -1.98 to 1.46 | -0.73 to 0.61 | -0.73 to 0.61 |
| $O_S + O_{ad}$ | -3.42 | -2.03 | -2.10 | -0.20 |
| $V_S + O_{ad}$ | 0.91 | 2.31 | 0.98 | 2.88 |

**Table S-I**. Formation energies of various point defects in ML-MoS$_2$ in Mo- and S-rich conditions at 0 K and 1023 K.

**IV) Determination of the electrostatic correction term for charged defect calculations:**

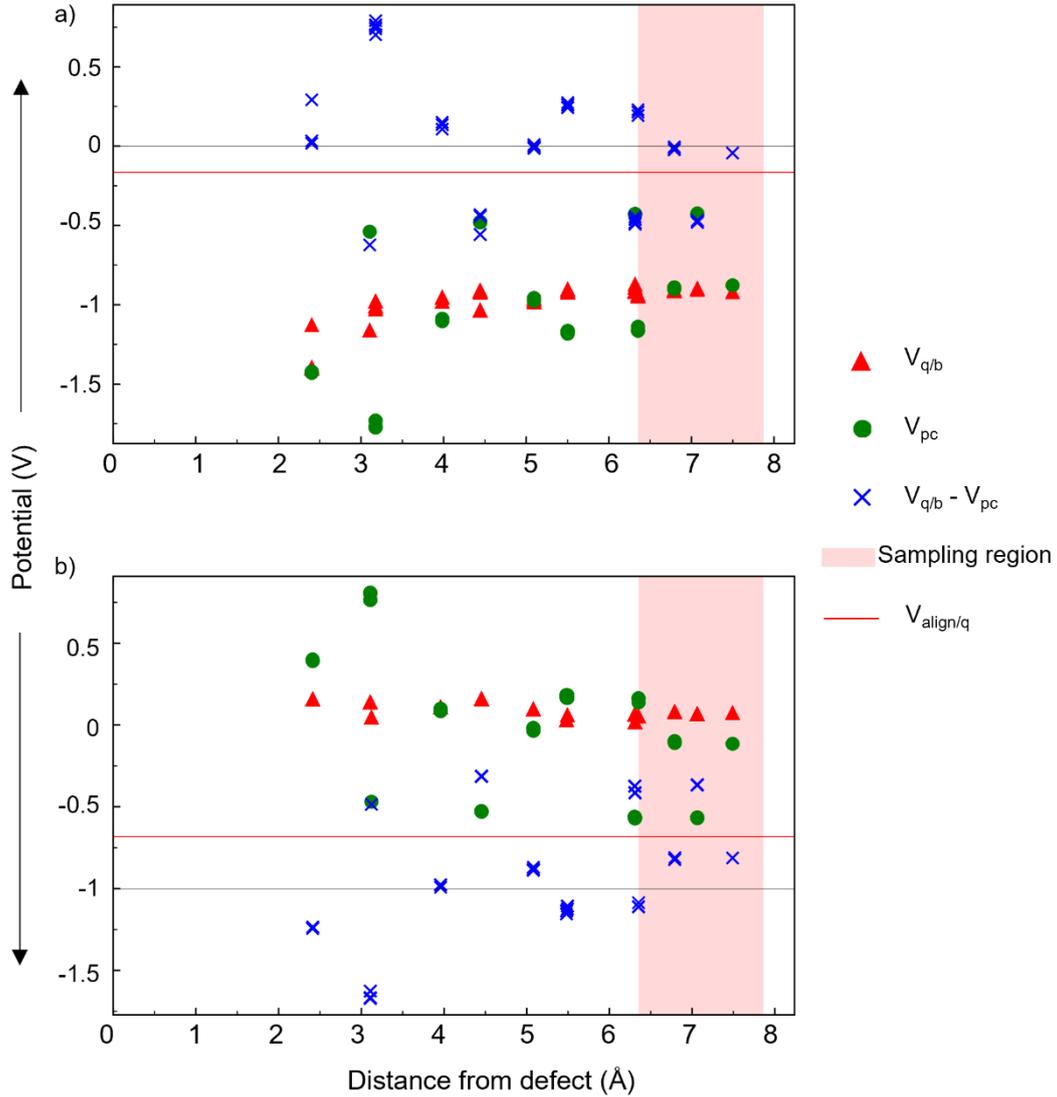

**Figure S-V.** Schematic of the potential alignment corrections, using the scheme of Kumagai and Oba,[5] in (a) negatively charged $V_S$ (q = -1) and (b) positively charged $V_S$ (q = +1). $V_{q/b}$ is the difference in density functional theory (DFT) calculated electrostatic potential between pristine and defective supercells, $V_{pc}$ is the model point-charge potential. $V_{q/b} - V_{pc}$ is the difference between the DFT-calculated potential difference and the point charge potential. $V_{align/q}$ reflects the average of $V_{q/b} - V_{pc}$, over the sampling region, which is represented by the shaded region (i.e., the region outside the Wigner-Seitz sphere around the defect). For a negatively charged $V_S$, the potential alignment term is given by $-q \times V_{align/q}$ = 0.16 eV with a net electrostatic correction ($E_{corr}$) of -0.12 eV and for a positively charged $V_S$, the alignment term is +0.32 eV with an $E_{corr}$ of -0.28 eV. We used the python charged defect toolkit (PyCDT)[6] for post-processing of charged defects calculation and determining $E_{corr}$.

**V) Calculated electronic density of states (DOS) of ML-MoS$_2$ in the presence of various defects:**

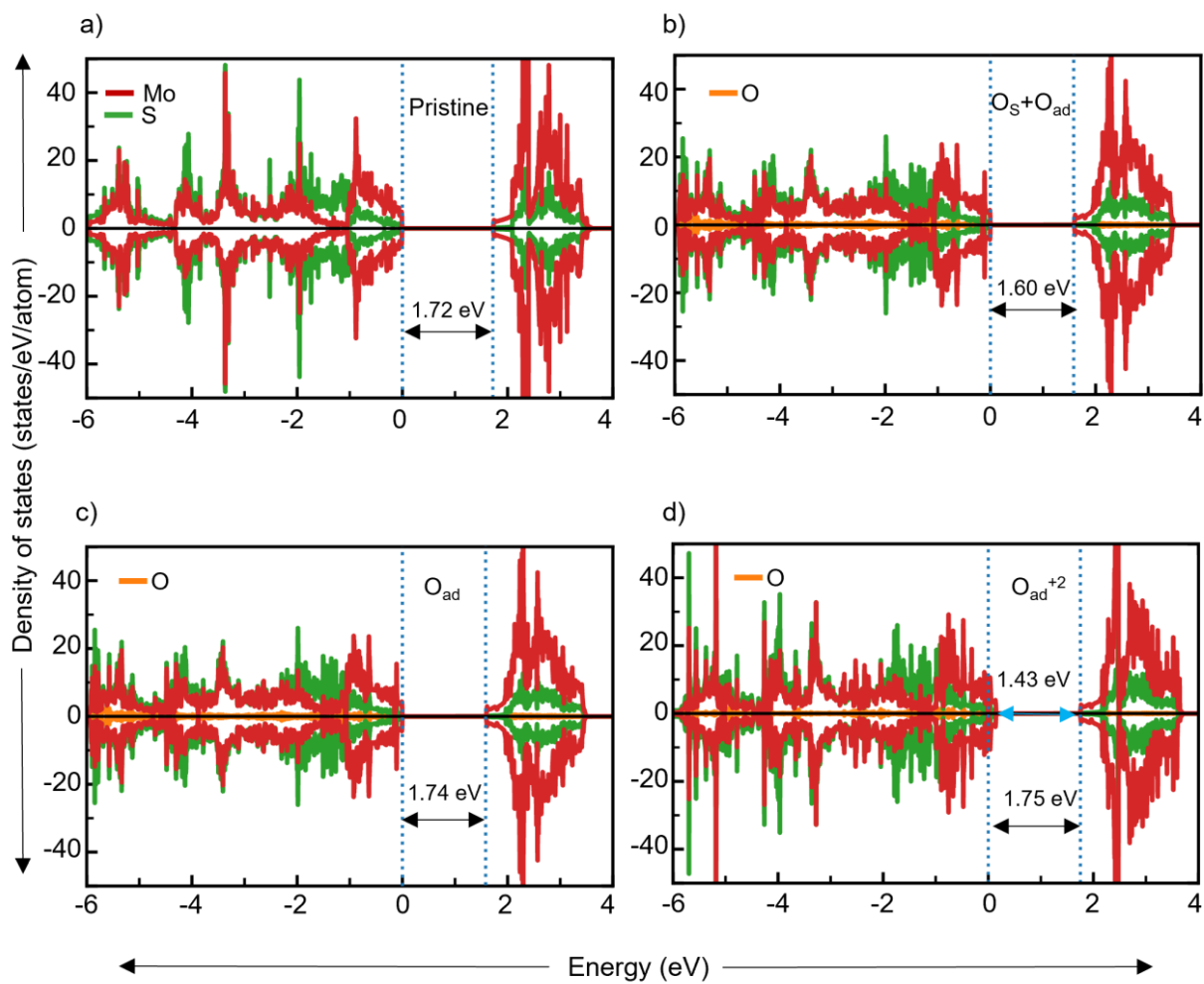

**Figure S-VI.** Density of states for a) Pristine ML-MoS$_2$ and MoS$_2$ with b) O$_S$ + O$_{ad}$, c) O$_{ad}$, and d) O$_{ad}^{+2}$. The dotted blue lines represent band edges. Mo-d, S-p, and O-p states are represented by red, green, and orange lines, respectively. Bandgap magnitudes are indicated by black arrows and defect levels by blue arrows.

## VI) Effect of defects and strain in the band structure and electronic DOS of MoS₂:

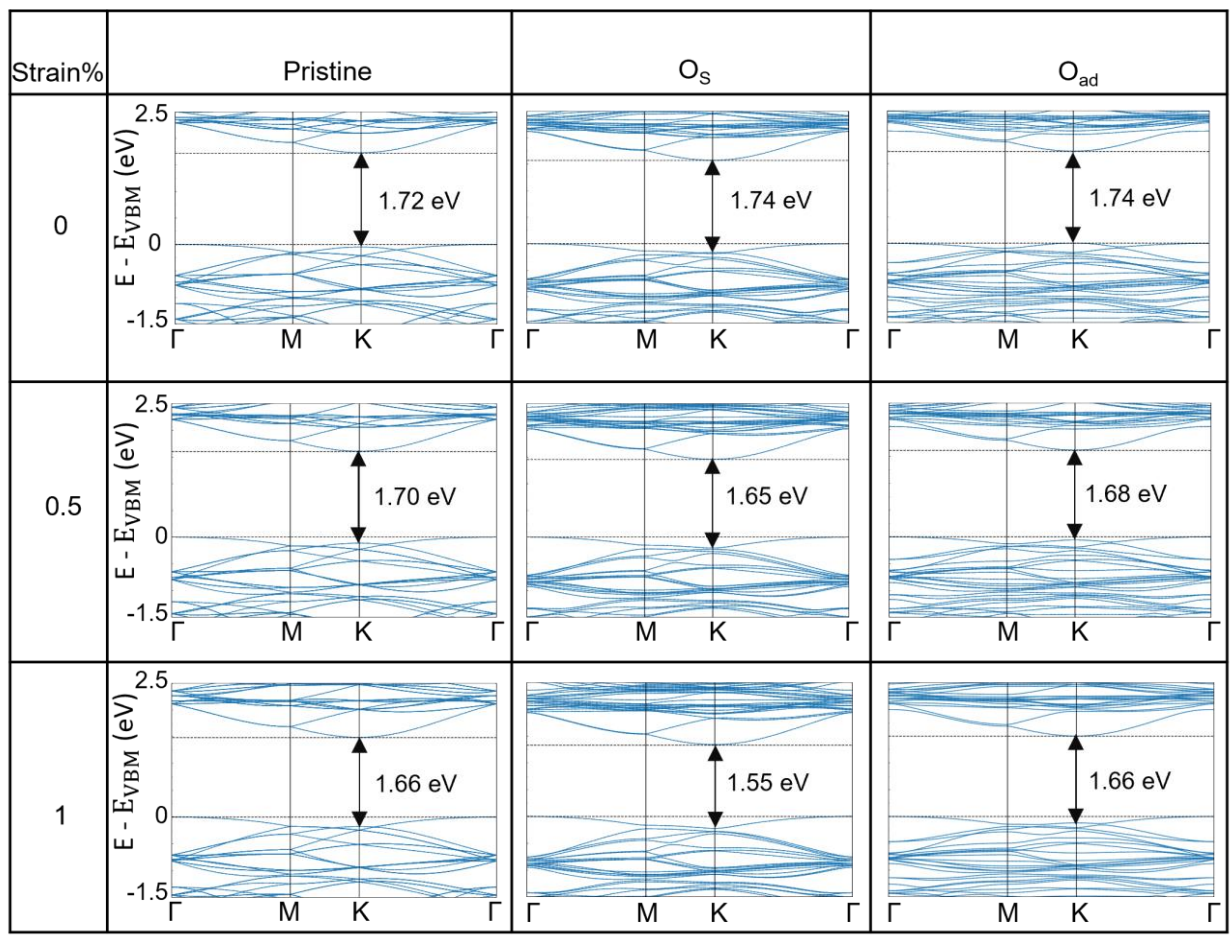

**Figure S-VII.** Band structures for Pristine ML-MoS$_2$, with O$_S$ and O$_{ad}$ at different applied strains. Dotted black lines indicate band edges. Black arrows indicate K-K bandgap.

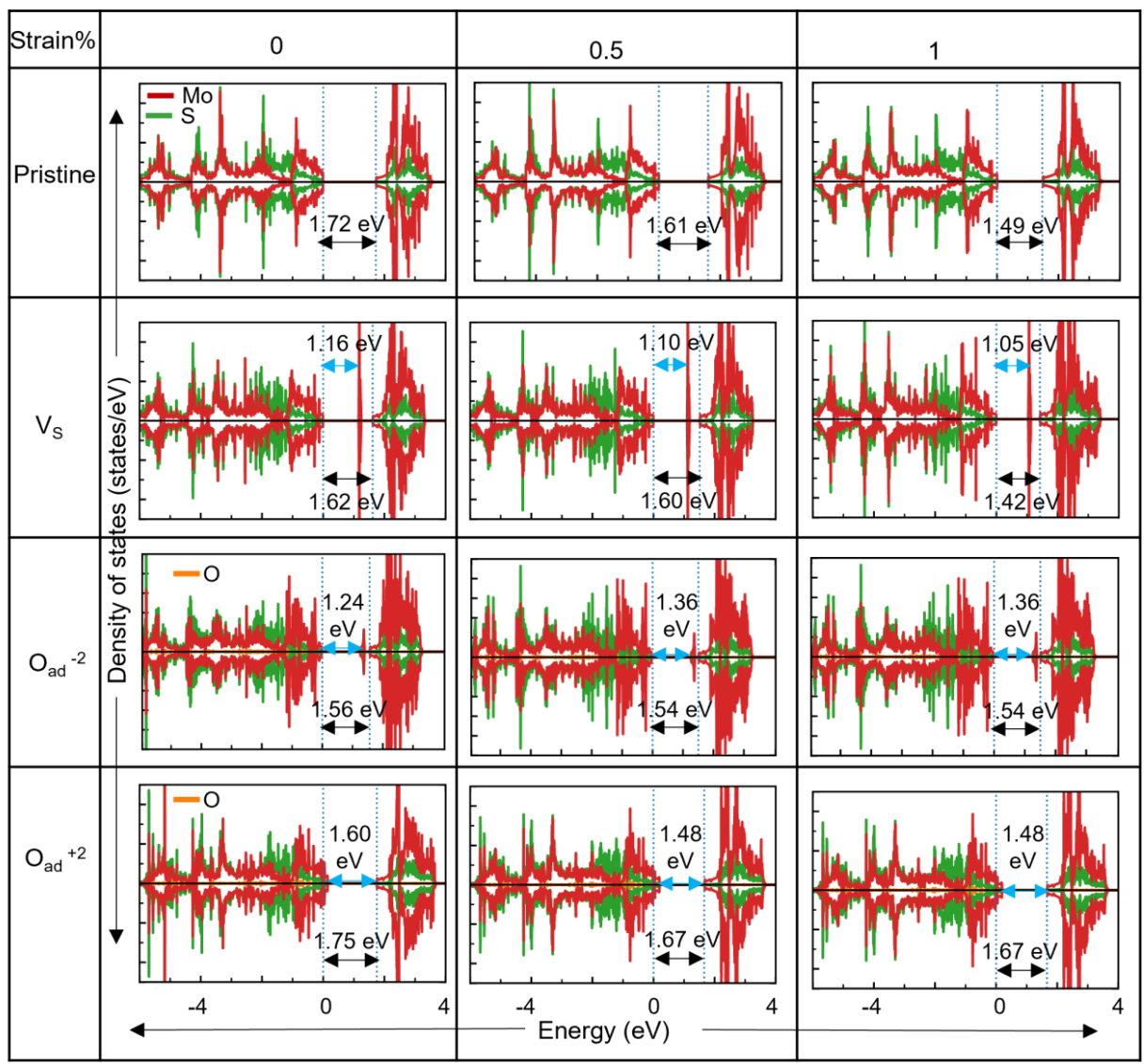

**Figure S-VIII.** DOS for Pristine ML-MoS$_2$, and MoS$_2$ with V$_S$, O$_{ad}^{-2}$ and O$_{ad}^{+2}$ at different degrees of applied strain. Dotted lines indicate band edges. Bandgap magnitudes (black arrows) are indicated along with defect levels (blue arrows) for V$_S$, O$_{ad}^{-2}$ and O$_{ad}^{+2}$.

## VII) Power dependent photoluminescence spectroscopy of hBN covered and encapsulated samples:

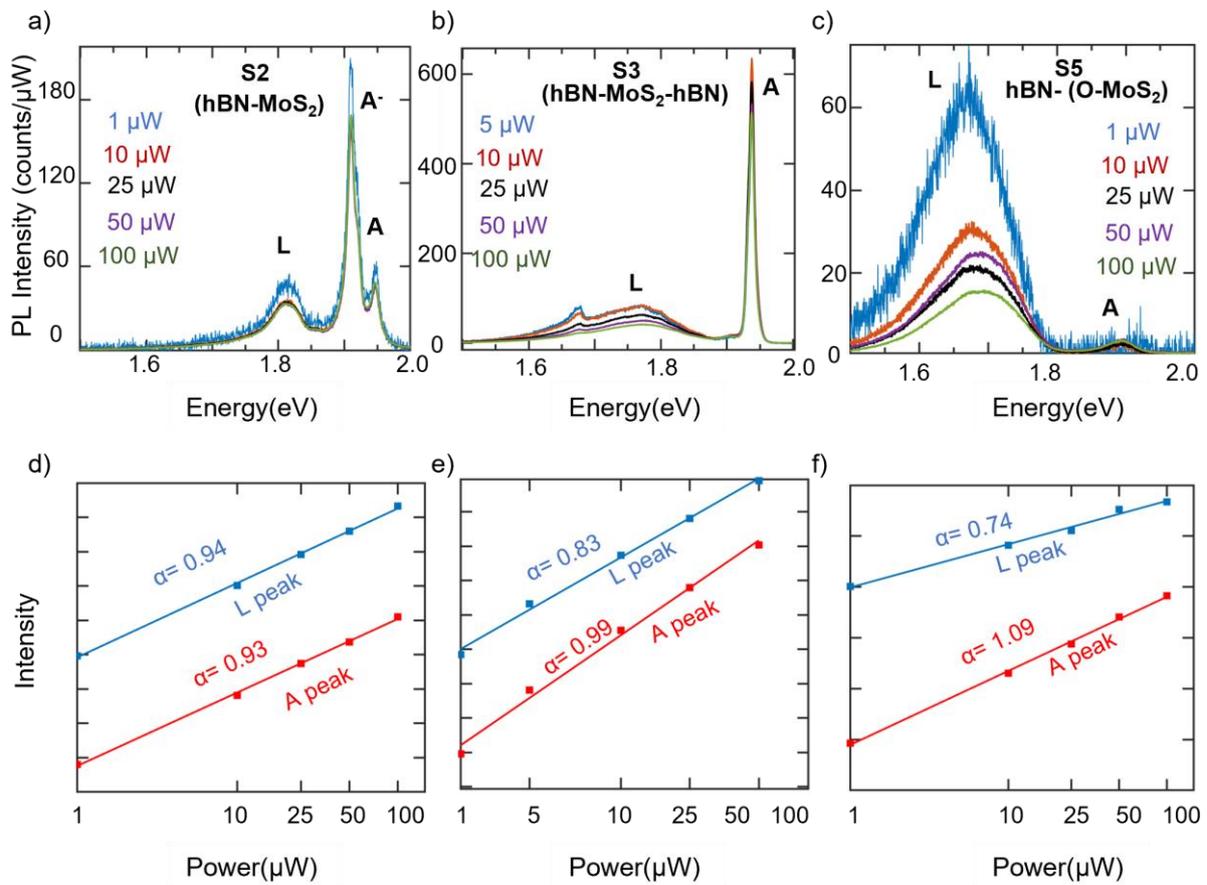

**Figure S-IX**. Power dependent photoluminescence (PL) spectra of a) S2 b) S3 c) S5 and log intensity v/s log power plots for d) S2 e) S3 and f) S5. All measurements were taken at 4K.

Figure S-IX shows normalized power dependent spectra of samples from S2, S3 and S5. All data is normalized to respective laser power values. The L peak behaves similarly in S3 and S5, but differently in S2 which has a nearly linear behavior. This can also be seen from the slope values in Figure S-IX d-f and inset of Figure 5a and 5b. The similarity in the peak behavior indicates similar origin of luminescence across the samples, but modified defect densities.

## VIII) Temperature dependent PL spectra of hBN covered and encapsulated samples:

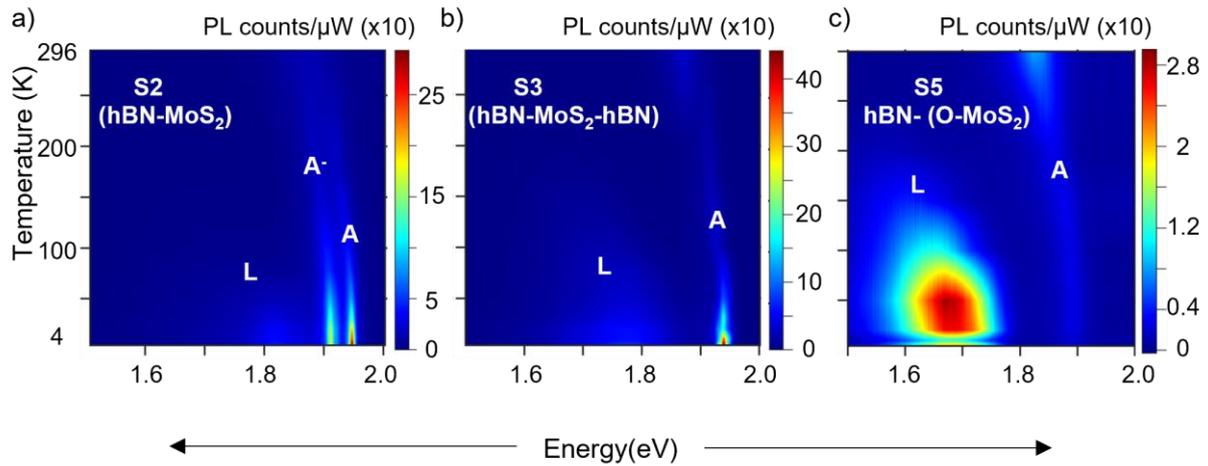

**Figure S-X**. Surface plots showing evolution of PL spectra with temperature for a) S2 b) S3 c) S5. Measurements were taken at 100 $\mu$W for S2 and S3 (same as S1 in Main text Figure 5), whereas 25 $\mu$W was used for S5 (same as S4 in Main text Figure 5).

Temperature dependent PL spectra of samples S2, S3 and S5 are shown in Figure S-X. For samples from S1-S3, both the A-excitonic peak and L-peak increases intensity as the temperature decreases with L-peak starting to appear at T~150 K. In S2 and S3, the L-peak intensity is comparatively less, as was observed in previous figures. S5 retains the behavior of S4 with A-exciton PL intensity decreasing with temperature, while L-peak intensity increases.

## IX) Comparison of room temperature Raman spectra:

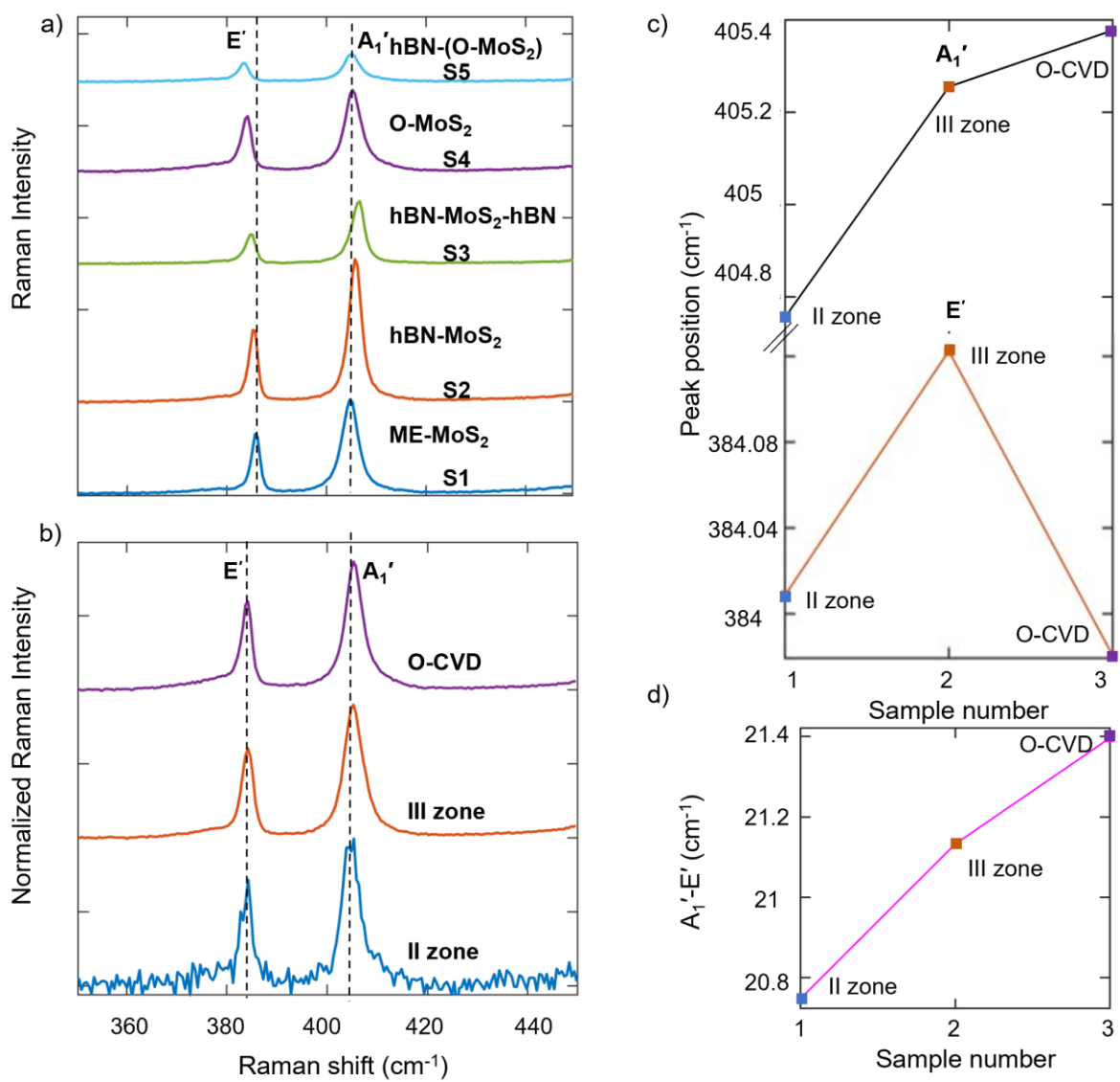

**Figure S-XI**. a) Background-subtracted Raman spectra of samples S1-S5. All data were taken at same acquisition conditions. The changes in the intensity of the peaks could be due to the presence of hBN. b) Comparison of Raman spectra c) Peak positions of $A_1'$ and $E'$ vibrational modes in Raman spectra and d) Difference between $A_1'$ and $E'$ peaks in Raman spectra in samples synthesized at different conditions using CVD. Raman intensity values are background subtracted and normalized to respective $A_1'$ peak intensity.

Figure S-XI (a) shows the background-subtracted Raman spectra of samples S1-S5, to show changed intensity for different samples. Background was calculated by averaging data in the spectral range away (340-360 cm$^{-1}$ range) from main Raman peaks, and then subtracted from raw Raman spectra. Figure S-

XI (b-d) compares the results of Raman spectroscopy on samples synthesized at different experimental conditions using CVD. In II zone CVD, the substrate is kept horizontally above the $MoO_3$ precursor boat while in III zone and O-CVD the substrate is kept vertically in a separate zone at a different temperature than the $MoO_3$ zone. No oxygen was used in samples other than O-CVD. The E′ peak doesn't seem to have evident shift confirming that the peak shift observed in Figure 6 is due to thermal strain. At the same time, $A_1$′ peak increases slightly indicating the possibility of increased doping in O-CVD sample due to the presence of oxygen.

## XI) XPS spectra of ML- MoS$_2$ synthesized using O-CVD and mechanical exfoliation:

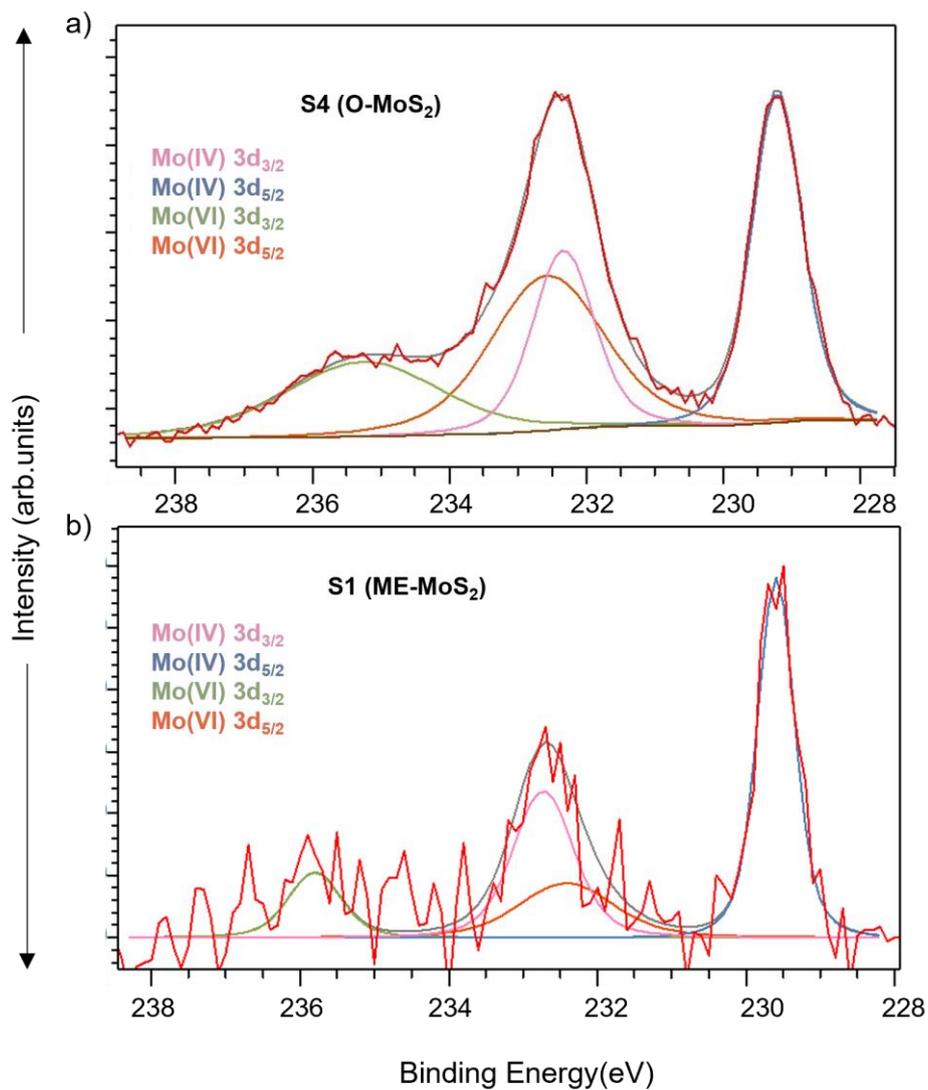

**Figure S-XII.** X-ray photoelectron spectroscopy (XPS) spectra of O-CVD and mechanically exfoliated ML-MoS$_2$ samples. a) Mo3d scan of ML MoS$_2$ synthesized using O-CVD b) Mo3d scan of ML MoS$_2$ obtained using mechanical exfoliation.

| Name | Position | | FWHM | | Weight percentage | |
|---|---|---|---|---|---|---|
| | S4 | S1 | S4 | S1 | S4 | S1 |
| Mo IV 3d 3/2 | 232.34 | 232.73 | 1.07 | 0.96 | 19.48 | 28.91 |
| Mo IV 3d 5/2 | 229.21 | 229.60 | 0.87 | 0.59 | 29.22 | 43.37 |
| Mo VI 3d 3/2 | 235.26 | 235.80 | 2.71 | 0.84 | 20.52 | 11.09 |
| Mo VI 3d 5/2 | 232.57 | 232.42 | 1.95 | 1.50 | 30.78 | 16.63 |

**Table S-II**. Fitted values from XPS spectra of S1 and S4.

To understand the chemical environment of the samples, including the presence of chemisorbed oxygen in the O-CVD sample and finding its concentration, finding the stoichiometry, and comparing the composition of different samples, XPS was performed (Figure S-XII). We observed that apart from $MoS_2$ peaks: $Mo^{4+}$ $3d_{3/2}$, $Mo^{4+}$ $3d_{5/2}$, there were two other peaks corresponding to $Mo^{6+}3d_{3/2}$ and $Mo^{6+}$ $3d_{5/2}$ indicating the presence of Mo-O bonds in both samples. The peak positions, FWHM and weight percentages are listed in Table S-II. Interestingly, the $Mo^{6+}$ peak is red shifted by 0.54 eV ($3d_{3/2}$) while $Mo^{4+}$ peaks are red shifted by 0.39 eV in the CVD sample. This might be indicative of lower n-type doping in the O-CVD sample.[9] The atomic percentage of $Mo^{4+}$ peaks combined is 48.7 and 72.28 for O-$MoS_2$ and exfoliated samples respectively and the percentage of $Mo^{6+}$ peaks are 51.3 and 27.72 respectively. The values obtained for $Mo^{6+}$ $3d_{5/2}$ for exfoliated samples is not considered for comparison because of the noisy nature of the data. The increased percentage of $Mo^{6+}$ peaks indicate increase in Mo-O bonds in the O-CVD sample. The Mo-O bonds in the exfoliated sample could be from physisorbed oxygen. Even though the exfoliated sample data is not of very high quality because of difficulty in performing XPS on an isolated 30 μm flake, the obtained values suggest different origins of oxygen (peak shift, difference in FWHM, changes in atomic percentage) confirming the presence of chemisorbed oxygen in the O-CVD sample.[9,10]

**XI) Room temperature PL spectra from multiple flakes on the O-CVD sample:**

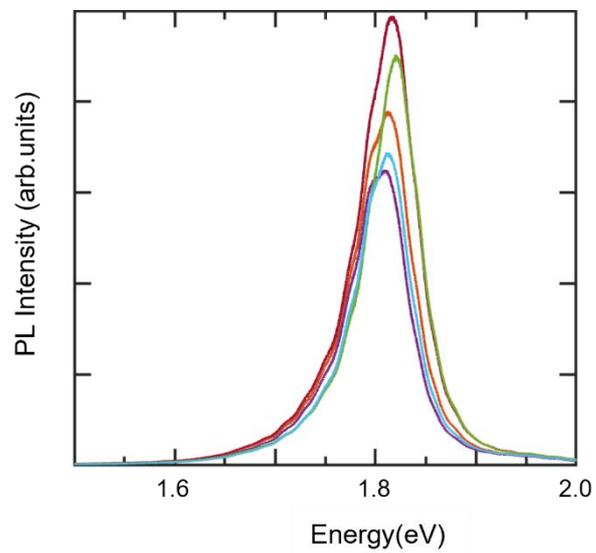

**Figure S-XIII**. Room temperature PL spectra collected from different flakes on O-CVD sample showing consistency of the obtained PL counts across the synthesized sample.

## XII) PL spectroscopy of O-CVD MoS$_2$ after hBN covering and annealing in glovebox:

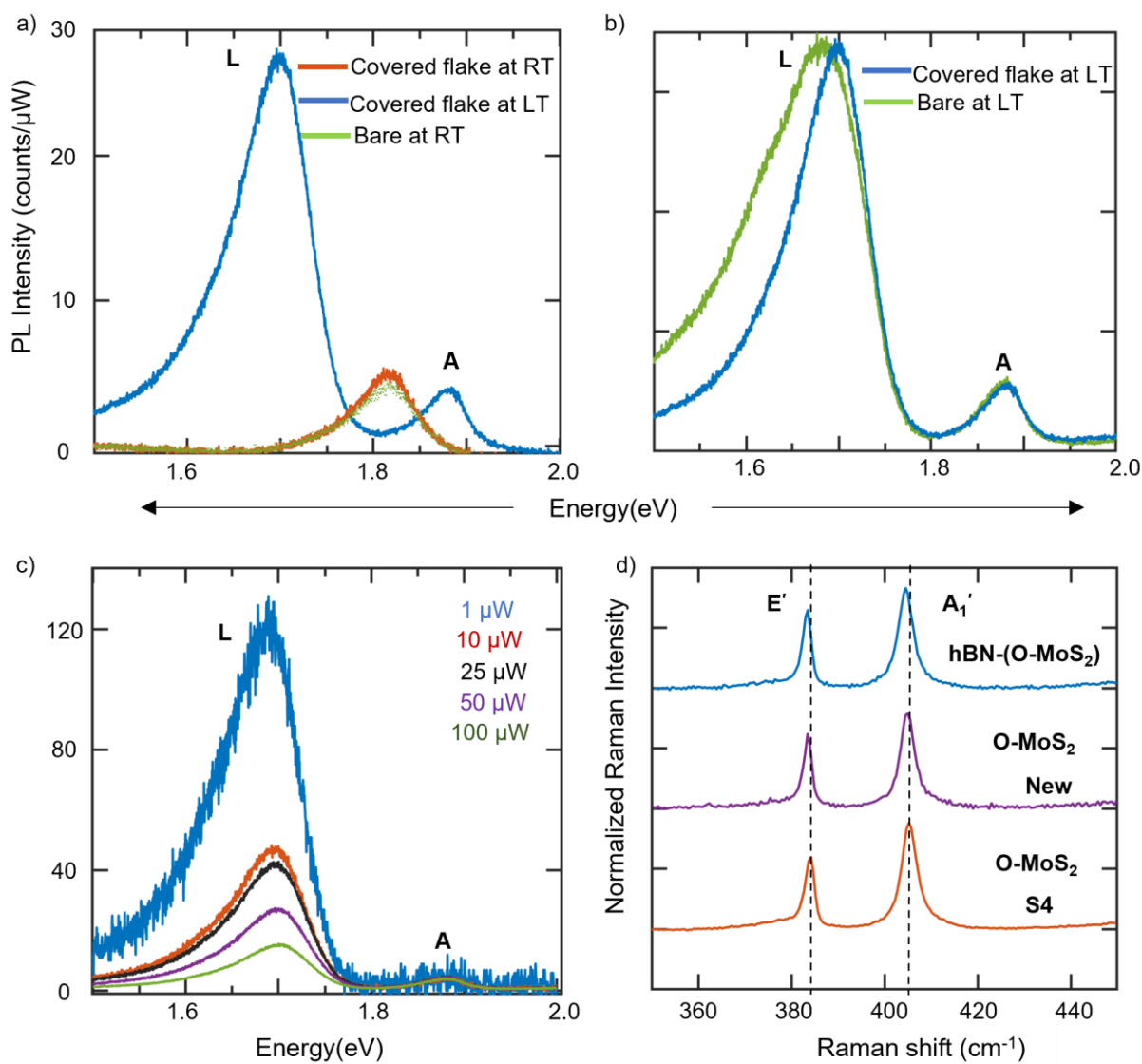

**Figure S-XIV**. a) Comparison of PL spectra at RT and LT before and after hBN covering b) Comparison of LT PL spectra of bare and covered flake (after the covering process). c) Power dependent PL spectra and c) Comparison of RT Raman spectra of hBN covered O-CVD sample annealed in glovebox.

Figure S-XIV a show the PL spectra of O-CVD sample after hBN covering and annealing was done in glovebox. We observed that similar to S5, the exciton intensity did not decrease after covering in this sample as well. This confirms the stability of the properties in the O-CVD sample despite the annealing conditions. We also observe an additional ~16 meV shift of the A exciton peak to the lower energy in comparison to S4 and S5, and a decrease in the FWHM of the L-peak in covered flake in comparison to the bare flake in the same sample (Figure S-XIV b). The decrease in FWHM indicates improvement

in the quality of the sample after hBN covering. This could be because of the elimination of hydrocarbon complexes in the covered material as discussed before. The power dependent spectra (Figure S-XIV c) show saturation of L peak with power suggesting similar defect population in the sample as S4 and S5. Figure S-XIV d shows comparison of Raman spectra of bare and covered flake with S4. There is a slight shift of E′ peak and $A_1'$ peak. The peak difference of the new sample is nearly the same as S4.

**XIII) Origin of L peak in exfoliated and O-CVD MoS$_2$ samples:**

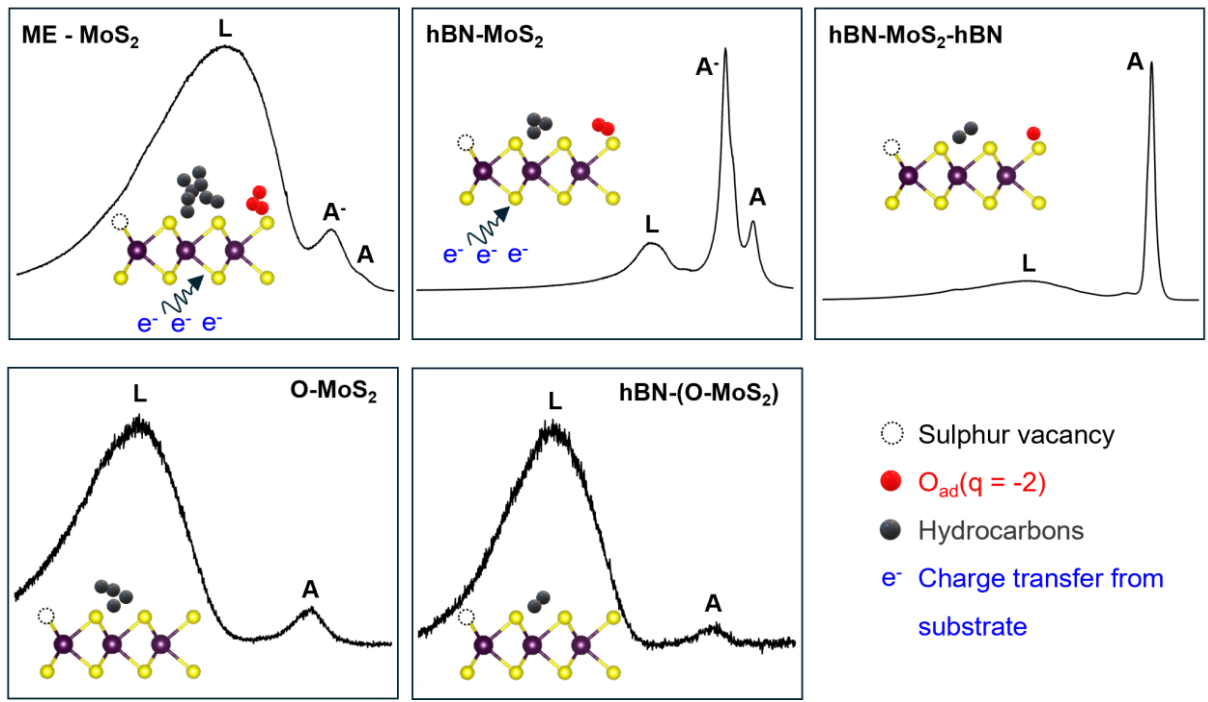

**Figure S-XV:** Schematics showing the factors contributing to different peaks in the LT PL spectra of samples S1-S5. The variations in the peak intensities can be directly related to the variations in type and amount of defects present in the samples. The defects contributing to L peak in exfoliated sample are sulphur vacancies, oxygen adatoms and hydrocarbon complexes and in O-CVD sample are sulphur vacancies and hydrocarbon complexes since charged oxygen adatom formation is less likely at the growth temperature. The importance of hBN covering and encapsulation in reducing adatoms and hydrocarbon complexes are clearly visualized. These observations are supported by results from formation energy, DOS and band structure calculations.

**XIV) Fitted parameters of L-peak and A-peak from LT PL spectra of samples from S1-S5:**

| Sample | Peak position for A-exciton (A-trion), meV | Peak position (L-peak) | FWHM (A-exciton) (meV) | Integrated intensity (A-exciton) (counts) | Integrated intensity (L-peak) (counts) |
|---|---|---|---|---|---|
| S1 | 1948 (1916) | 1797 | 50 | 1.02e5 | 8.97e6 |
| S2 | 1945 (1912) | 1818 | 20 | 1.93e5 | 8.22e5 |
| S3 | 1939 (1904) | 1766 | 15.6 | 1.54e6 | 3.64e6 |
| S4 | 1891 | 1672 | 60 | 6.18e4 | 1.28e6 |
| S5 | 1890 | 1679 | 62 | 5.75e4 | 7.43e5 |

**Table S-III**. Extracted values of L-peak and A-peak from LT PL spectra of samples S1-S5.

Table S-III shows the extracted values of peak position, FWHM and integrated intensities of L-peak and A-peak in the LT PL spectra of samples S1-S5. All the data used for fitting were taken at 100 μW laser power. The L-peak in S1, S4 and S5 were fitted using three peaks and the integrated intensity is the sum of all the peaks. The reported peak position of L-peak in S1, S4 and S5 are the maximum intensity point of the broad peak. In S2 and S3, since L peak intensity was much less than trion and exciton peaks, only one peak was used for fitting. Trion peak was indistinguishable from exciton peak in S3, S4 and S5, and thus were fitted using a single peak while it was fitted using separate peaks in S1 and S2.